\documentclass{rspublic}
 
\usepackage{graphicx}

\begin{document}
\title[MHD evolution of magnetic skeletons]{Magnetohydrodynamic evolution of magnetic skeletons}
\author[A.L. Haynes and others]{Andrew L. Haynes$^1$, Clare E. Parnell$^1$, Klaus Galsgaard$^2$ and Eric R. Priest$^1$ }
\affiliation{$^1$School of Mathematics and Statistics, University of St. Andrews, St.
  Andrews KY16 9SS, UK and
  $^2$Niels Bohr Institute, Julie Maries vej
  31, 2100 Copenhagen \O, Denmark}
\label{firstpage}
\maketitle
\begin{abstract}{Magnetohydrodynamics (MHD), Sun, magnetic fields}
The heating of the solar corona is likely to be due to reconnection of the highly complex magnetic field that threads throughout its volume. We have run a numerical experiment of an elementary interaction between the magnetic field of two photospheric sources in an overlying field that represents a fundamental building block of the coronal heating process. The key to explaining where, how and how much energy is released during such an interaction is to calculate the resulting evolution of the magnetic skeleton. A skeleton is essentially the web of magnetic flux surfaces (called separatrix surfaces) that separate the coronal volume into topologically distinct parts.  For the first time the skeleton of the magnetic
field in a 3D numerical MHD experiment is calculated and carefully analysed, as are the ways in which it bifurcates into different topologies. A change in topology normally changes the number of magnetic reconnection sites.

In our experiment, the magnetic field evolves through a total of six distinct topologies. Initially, no magnetic flux joins the two sources.  Then a new type of bifurcation, called a \emph{global double-separator
bifurcation}, takes place: this bifurcation is likely to be one of the main ways in which new separators  are created in the corona (separators are field lines at which 3D reconnection takes place). 
This is the first of five bifurcations in which the skeleton becomes progressively more complex before simplifying. Surprisingly, for such a simple initial state, at the peak of complexity there are five separators and eight flux domains present. 
\end{abstract}

\section{Introduction}
 
The solar corona, the upper atmosphere of the Sun, is known to be a low-density
plasma with a temperature greater than $1 \times 10^{6}$ K, while the surface
of the Sun - the photosphere - has a temperature of about 6000K. The solar
atmosphere consists of active areas around sunspots, and regions of lower
activity known as the quiet Sun. The quiet-Sun corona is, however, far from
quiet and consists of magnetically open coronal holes, many types of coronal
loop and dots of emission with a range of scales, such as microflares and 
X-ray 
bright points. Although the heating of X-ray bright points, and most probably 
microflares, has been shown to be produced by magnetic
reconnection (Priest et al. 1994; Parnell at al. 1994a,b), the heating of 
coronal holes and coronal loops remains an unsolved mystery.
 
It has long been known that the heating mechanisms depend on the
magnetic fields that are present over the surface and throughout the atmosphere
of the Sun. For many years these mechanisms were generally divided into two
categories: magnetic waves and magnetic reconnection. Although low-frequency
waves have now been observed in the corona (e.g., Nakariakov et al. 1999; De Moortel et al. 2002) and are important for
coronal seismology (Roberts 1984; De Moortel 2005; Nakariakov \& Verwichte 2005), they do not carry enough energy to heat
the corona and so reconnection is now regarded as the most likely mechanism.
 
Various reconnection mechanisms have been proposed, including those that require the presence of nulls (e.g., Lau \& Finn 1990; Priest \& Titov 1996; Longcope 2001; Priest et al. 2003) and those that can occur in the absence of nulls (e.g., Parker 1972; Priest \& D{\'e}moulin 1995; Birn et al. 1998).

The quiet-Sun photosphere is covered by small magnetic
fragments, which are driven across the surface by photospheric flows. New flux appears through a process called
emergence which creates two opposite polarity fragments forming an ephemeral region 
(Harvey \& Martin 1973). Flux is later removed through the process of cancellation when two 
opposite-polarity fragments collide and mutually lose flux (Martin et al. 1985; Harvey 1985). Cancellation and emergence lead to reconnection that can power an X-ray bright point as explained by Priest et al. (1994), Parnell et al. (1994a,b) and von Rekowski et al. (2006a,b).
However, the simple moving around of magnetic fragments in the photosphere 
also causes magnetic reconnection in the atmosphere (Longcope 1998; Galsgaard et al. 2000; Parnell \& Galsgaard 2004; Galsgaard \& Parnell 2005). The ceaseless movement of photospheric fragments in 
the quiet Sun drives a wealth of reconnection which can power X-ray bright 
points, microflares and nanoflares. Indeed, Close et al. (2004, 2005a) 
determined 
a conservative estimate that coronal magnetic fields are recycled ten 
times faster (in just 1.4 hrs) than photospheric magnetic fields, implying 
that there is considerably more reconnection occurring throughout the solar 
atmosphere than is driven by emergence and cancellation alone. 

In this paper, we use the term {\it fly-by} to describe the passing of nearby 
fragments where flux is conserved. 
Using a 3D MHD code, Galsgaard et al. (2000), Parnell \& Galsgaard (2004) and Galsgaard \& Parnell (2005) calculated the reconnection rate and energy release of a 
simple set of fly-by models. They considered two discrete source fragments 
of opposite polarity, situated in an
overlying field.  The sources were driven anti-parallel to each other by
advection flows in the photosphere at various angles to the overlying
field. The flux through the base was conserved throughout. The motions caused
the flux lobe from each source to be pulled under the flux lobe from the other
source. A twisted current sheet was formed between the two flux lobes as the
sources became connected. Later, new current sheets started near the ends of
the connected flux region and extended along it as the sources were
disconnecting.
It was suggested that
the connection/linking of the sources is caused by {\it separator
reconnection} (see also Longcope 2001; Priest \& Titov 1996) in the twisted current sheet surrounding this separator, and
that the sources later disconnected through a process that they called 
separatrix-surface
reconnection. Furthermore, the rates of reconnection were found to be fast: up to 58\% of the instantaneous, perfect reconnection rate of the potential evolution. Although, the rates of reconnection were found to be almost a
 factor of two slower in the disconnecting phase than in the connecting phase. 
 
Our primary aim is to understand much better where and how reconnection
 is occurring in the solar corona by first understanding in detail just how it is occurring in the fly-by numerical experiments of Galsgaard and
Parnell. To shed light on this, we must first answer the following two 
fundamental questions in solar MHD: 

1. What is the topological structure
of the magnetic field (i.e. where are the possible reconnection sites) as separate 3D magnetic flux systems interact in the 
solar atmosphere, and 

2. What kind of bifurcation leads to a change of topology (i.e. how and why do the number of reconnection sites change)? 

\noindent As a background, we need first to say a little about magnetic topologies and 
about 3D magnetic reconnection.

First of all, the topological structure of a complex magnetic field is best
described in terms of its \emph{magnetic skeleton} defined by e.g. Bungey et al. (1996) and Priest et al. (1997) on the basis of earlier work and definitions by e.g. Greene (1988), Lau \& Finn (1990) and Priest \& Titov (1996). Such a skeleton includes:

  1. \emph{sources}, where the magnetic field
    enters or leaves the region being studied,

  2. \emph{null points} where the magnetic field strength vanishes,

  3. \emph{separatrix surfaces}, which are made up of
    field lines that extend to or from a null and form the
    boundary between topologically distinct domains (also known as $\Sigma$ or 
    fan surfaces),

  4. \emph{spines}, isolated field lines extending from a 
    null back to a sink, if the separatrix surface from the null goes to a 
    source, or vice versa (also known as $\gamma$ lines),

  5. \emph{separators}, which are the
    intersections of two separatrix surfaces. A separator connects two null
    points and represents the dividing line between four topologically distinct
    domains (also known as A-B lines).
 
Next we must define what a flux domain is. A \emph{flux domain} is a simply-connected volume of magnetic field lines that share the same start and end sources or \emph{source pair} and is bounded by a separatrix surface (Longcope \& Klapper 2002; Beveridge \& Longcope 2005 and Parnell 2007). Thus, in such a domain it is possible to continuously deform the field lines from one to another. This definition is related to the key feature that is generally used to find most flux domains, namely the connectivity of field lines, but it goes much further. If there were two separate paths linking the same source pair, each surrounded by their own separatrix surfaces, and if each of these volumes were simply connected then they would be classed as \emph{flux domains of a multiply-connected source pair}. The main point here is that knowing the number of connected source pairs is not the same as knowing the topological structure of the field. 

In general, it is assumed that most source pairs have multiplicity one. Although as soon as the magnetic field becomes complex, as it is on the Sun, this is not necessarily the case. Source pairs with multiplicity two were found by Close et al. (2005b) and Barnes et al. (2005) in potential magnetic field extrapolations of two different magnetograms. In the present experiment, we find source pairs with multiplicity up to three. 

The simple potential topologies associated with three or four flux sources and
the bifurcations between the topologies have been studied by
Brown \& Priest (1999), Beveridge et al. (2003) and Maclean et al. (2006). In particular, they
discovered \emph{local bifurcations}, in which the topology changes when pairs
of null points are created or destroyed, and \emph{global bifurcations}, in
which the number of null points is held fixed during a change of topology. A
comprehensive review of such topologies has recently been given by
Longcope (2005). 
Of particular interest to our numerical experiment is the 
Beveridge-Longcope equation (Beveridge \& Longcope 2005 and Parnell 2007), namely,
\begin{equation} \label{eq:mod-bev-long}
{\cal D} = \sum_{n} nD_n = X + S - N_c - 1,
\end{equation}
which relates the number of separators ($X$), sources ($S$) and coronal nulls 
($N_c$) to the number of domains (${\cal D} = \sum_n nD_n$), where $D_n$ 
denotes the number of source pairs with multiplicity $n$.

 In this paper, we compare our numerical MHD results with 
those for potential fields and discover that MHD evolution leads to a much 
richer and more complex set of topologies than evolution through a series of 
equi-potential states. 

The paper is divided into the following parts. The next section, \S2, is a
description of the potential and MHD models, while \S3 describes
the evolution of their skeletons. \S4 and \S5 highlight the nature of the
bifurcations and the development of multiply-connected domains. Finally, in the
last section, we discuss the implications of the results.

\section{Numerical Model} \label{sec:num}

The numerical model used here has the same set-up as in Galsgaard et al. (2000),
 Parnell \& Galsgaard (2004) and Galsgaard \& Parnell (2005). We give a summary here for completeness.

The setup comprises a Cartesian box of $128 \times 128 \times 65 $ grid
points with a scale of $1 \times 1 \times 1/4$. The normal components of the 
initial magnetic field are imposed on the base ($z = 0$).
They consist of a positive source, of radius $r_{0} = 0.065$ and maximum 
magnetic field strength $B_{0} = 0.85$, placed at $(\frac{1}{3},
\frac{1}{3}, 0)$ and a negative source with the same radius and maximum magnetic
field strength, but opposite polarity, placed at $(\frac{2}{3},
\frac{2}{3}, 0)$.  These sources have a cosine profile of the form $B_{z} =
B_{0}\left(1 + \cos ({\pi r}/{r_{0}}) \right)/2$, where $r < r_{0}$ is the
distance from the centre of the source. Elsewhere on the base, the normal
component of magnetic field is taken to be zero.  Otherwise, the box is
closed at the top and bottom and periodic in $x$ and $y$.  An overlying field, ${\bf B_{over}} = 0.12
B_{0}{\bf \hat{y}}$ is added to disconnect the two sources in the initial state.  For comparison we consider a potential model and a dynamical
MHD model which are described below.
\subsection{Potential Model} \label{sub:num-pot}

The potential magnetic field (${\bf B}_p$), by definition, satisfies
$\nabla\times{\bf B_p} = {\bf 0}$ and $\nabla.{\bf B_p} = 0$, so that it may be written as ${\bf B_p} = \nabla\phi$, where $\phi$
satisfies
\begin{equation} \label{eq:num-pot}
  \nabla^{2}\phi = 0.
\end{equation}
Using a multigrid method with a Gauss-Seidel smoother and the above boundary
conditions imposed on a unit box, we can solve equation~\ref{eq:num-pot} for $\phi$ at each time frame, from which we evaluate ${\bf B_p}$. The taller numerical box of height 1 for the potential field in comparison to the 1/4 for the MHD model makes essentially no difference to the results as all the magnetic field at height 1/4 and above is simply the horizontal overlying field.

\subsection{Dynamic MHD Model} \label{sub:num-mhd}

The dynamic MHD approach uses the non-dimensionalised, non-ideal MHD equations
in the form:
\begin{eqnarray*}
  \frac{\partial (\rho {\bf u})}{\partial t} & = & -\nabla.\left(\rho {\bf u}
  {\bf u} + {\bar{\bar\tau}}\right) - \nabla p + {\bf J} \times
  {\bf B}, \\
  \frac{\partial e}{\partial t} & = & -\nabla.(e{\bf u}) - p\nabla.{\bf u} -
  \frac{\rho(T - T_{0})}{t_\mathrm{cool}} + Q_\mathrm{visc} + Q_\mathrm{joule},
    \\ 
      \frac{\partial {\bf B}}{\partial t} & = & -\nabla \times {\bf E}, \hspace{2.75cm}
  \frac{\partial \rho}{\partial t} = -\nabla.(\rho {\bf u}), \\
  {\bf E} & = & -\left( {\bf u} \times {\bf B} \right) + \eta {\bf J}, \hspace{1.92cm} {\bf J}  =  \nabla \times {\bf B}, \\
  p & = & (\gamma - 1)e, \hspace{3.0cm} T = \frac{p}{\rho},
\end{eqnarray*}
where 
${\bf B}$, ${\bf u}$, $e$, ${\bf J}$, ${\bf E}$,  $p$, $\rho$, $T$, $T_{0}$,
$t$, $t_\mathrm{cool}$, $\eta$, $Q_\mathrm{joule}$, $Q_\mathrm{visc}$ and
${\bar{\bar\tau}}$ are the magnetic field, velocity, internal energy, current,
electric field, pressure, density, temperature, equilibrium temperature, time,
exponential cooling time, magnetic resistivity, joule dissipation, viscous
dissipation and viscous stress tensor, respectively.  An ideal gas with the
ratio of specific heats $\gamma = 5/3$ is assumed.  The equations have been
non-dimensionalised such that the permeability of free space ($\mu_{0}$) and 
the ratio of the gas constant over the mean molecular weight (${\tilde{R}}/{
\tilde{\mu}}$) both disappear from the equations.

Initially, we assume an equilibrium with a potential magnetic field in an
atmosphere with uniform density and pressure of 1/2 and 1/6, respectively.  
The magnetic field is calculated using the method described in 
\S2a using the two sources defined above in their initial state.  

These sources are advected rigidly by flows in strips that are just wider than
their corresponding sources, with the velocities ramped down along the edges of
these strips to zero.  The flows are accelerated from rest up to a constant
speed of $0.02$ of the peak Alfv\'en speed in approximately $0.6$ of an 
Alfv\'en time. The
positive source is driven along the base in the ${\bf \hat{x}}$ direction, and
the negative source driven in the $-{\bf \hat{x}}$ direction.

The equations are solved using a third-order predictor-corrector method
with a sixth-order method used to calculate the spatial derivatives.  As
staggered grids are applied, a fifth-order method is used for spatial
interpolation. Viscosity and magnetic resistivity are handled using a
fourth-order method combined
with a discontinuous capture mechanism to provide the highest possible
spatial resolution for the given numerical resolution.
Diffusion, and therefore magnetic reconnection, is anticipated to take place in current sheets that have a very narrow, fine-scale structure. In 
codes using uniform or anomalous current dependent resistivity 
(i.e. $\eta\propto |{\bf j}|$) the resistivity ($\eta$) has a $k^2$ dependence, where $k$ is the wave number. This means that significant diffusion can occur even in well resolved current structures.
The advantage of the hyper-resistivity used here is that a higher $k$ dependence ($k^4$) is achieved implying that the dissipation increases very rapidly as the wavelength decreases localising the diffusion more efficiently to short length-scales. This is equivalent to having finer-scale current sheets, but with the same numerical grid resolution. To achieve this the resistive and viscous dissipation are dependent on three factors, namely, (i) compression regions (for the resistivity this compression is only relevant when it acts perpendicular to the magnetic field vector), (ii) advection speed (plasma velocity) and (iii) the typical propagation speed of information (fast-mode/sound speed). More details of how these factors influence the viscosity and resistivity and a description of the code can be found at {http://www.astro.ku.dk/$\sim$kg/}.

Further particulars of
the experiment can be found in Parnell \& Galsgaard (2004). All the times in this paper are 
given in terms of the initial Alfv{\'e}n time of the experiment.

\section{Evolution of Magnetic Skeleton} \label{sec:evol}

The corona has often been assumed to be near potential and many theoretical
treatments have, for simplicity, assumed potential fields, the minimum-energy
magnetic field for any normal field component prescribed. Potential fields may evolve through a series of equi-potential states by using perfect and instantaneous reconnection. It is
useful, therefore, to compare the differences between the skeleton of our
dynamic MHD magnetic fields with those that would arise from a potential field.

For this, we classify the four different types of flux domain that are found in
our experiment by way of their flux connectivity (or source pairs).  Flux in the
\textit{overlying} domain is not connected to either source (i.e. it
connects one side boundary to another side boundary).  The \textit{positive
open}(\textit{negative open}) domain contains flux that connects the positive(negative) source to a
side boundary.  Flux in
the \textit{closed} domain connects the positive source to the
negative source.  The boundaries of these flux domains define the skeleton. We
discuss the evolution of skeletons of the potential and dynamic MHD models in
turn.

\subsection{Potential Skeleton} \label{sub:evol-pot}

The potential field is calculated using the method described in
\S2a for each time step, from which we determine the
magnetic skeleton.  We visualise the skeleton in 3D (figure~\ref{fig:pot-phases}, col 1). The footprint of the skeleton represents the
intersection of the 3D skeleton with the photosphere
(figure~\ref{fig:pot-phases}, col 2). All null points found are photospheric and the footprint is calculated by tracing the 
separatrix field lines that lie in the photospheric source plane from 
these null points. The spines are not shown.
A cross-section of the skeleton at $y=0.5$, which lies midway between both of 
the sources and the two null points, is also determined (figure~\ref{fig:pot-phases}, col 2).  Both 
separatrix surfaces and all of the separators intersect this plane. 

By considering the evolution of the field, we group contiguous time
frames into phases with the same configuration of flux domains.  Four phases
are found (see table \ref{tab:phases-pot}): P1 - initial open, P2 - closing,
P3 - reopening and P4 - final open. We now describe 
these four phases in turn.
\subsubsection*{Phase P1 ($0.00<t<1.47$): Initial Open}
\begin{figure}[ht]
  \begin{tabular}{ccc}
    \resizebox{.30\hsize}{.22\hsize}{%
        \includegraphics{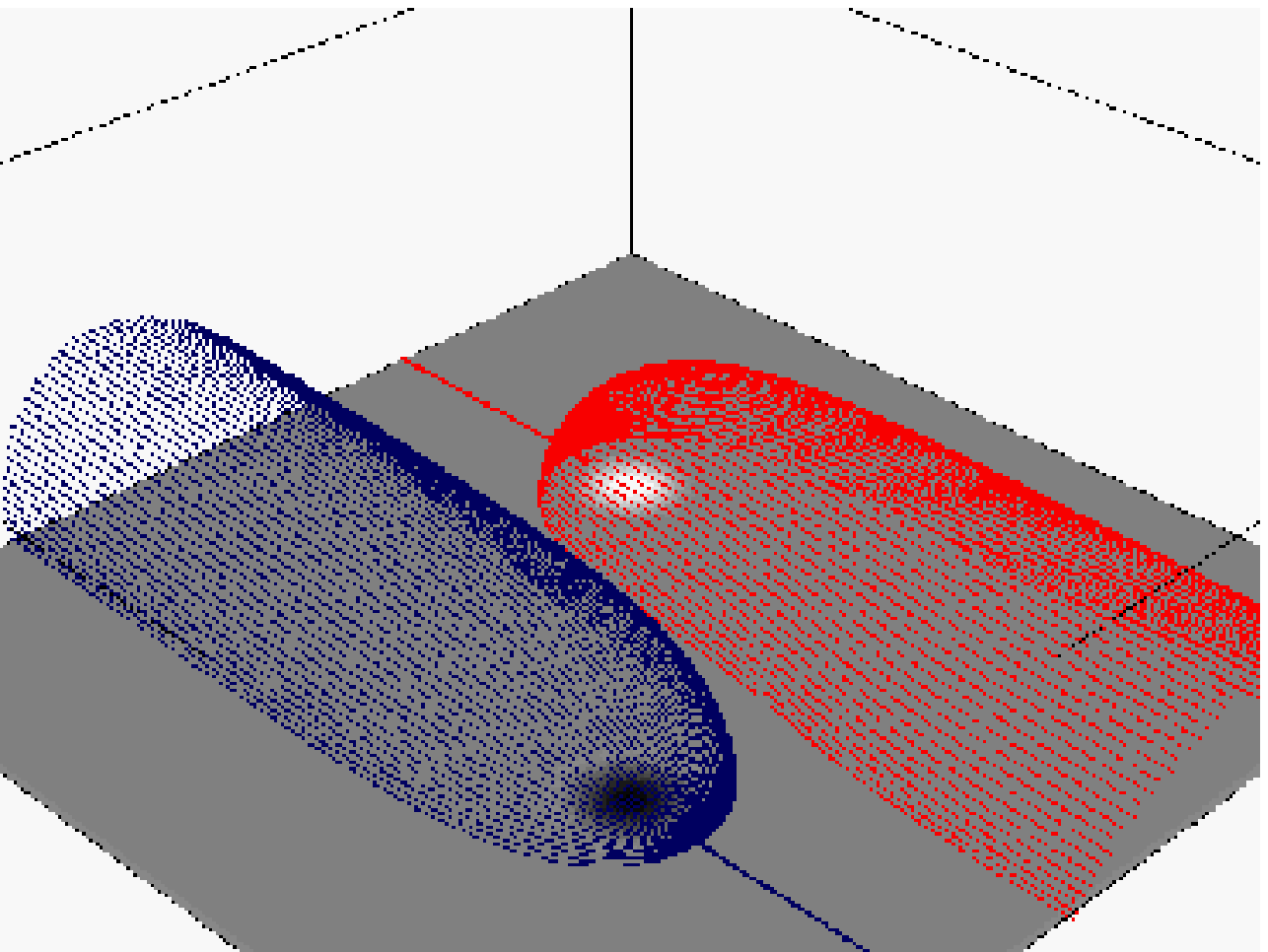}} &
      \resizebox{.30\hsize}{.22\hsize}{%
        \includegraphics{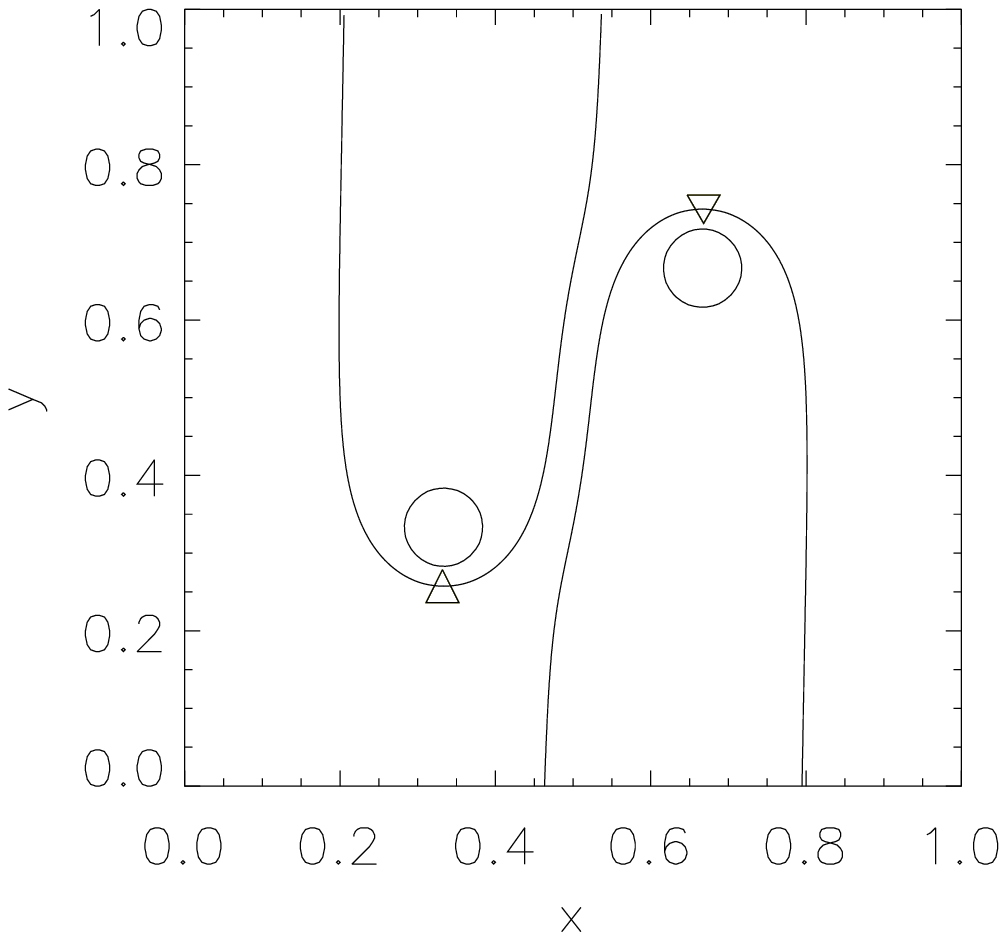}} &
      \resizebox{.30\hsize}{.22\hsize}{%
        \includegraphics{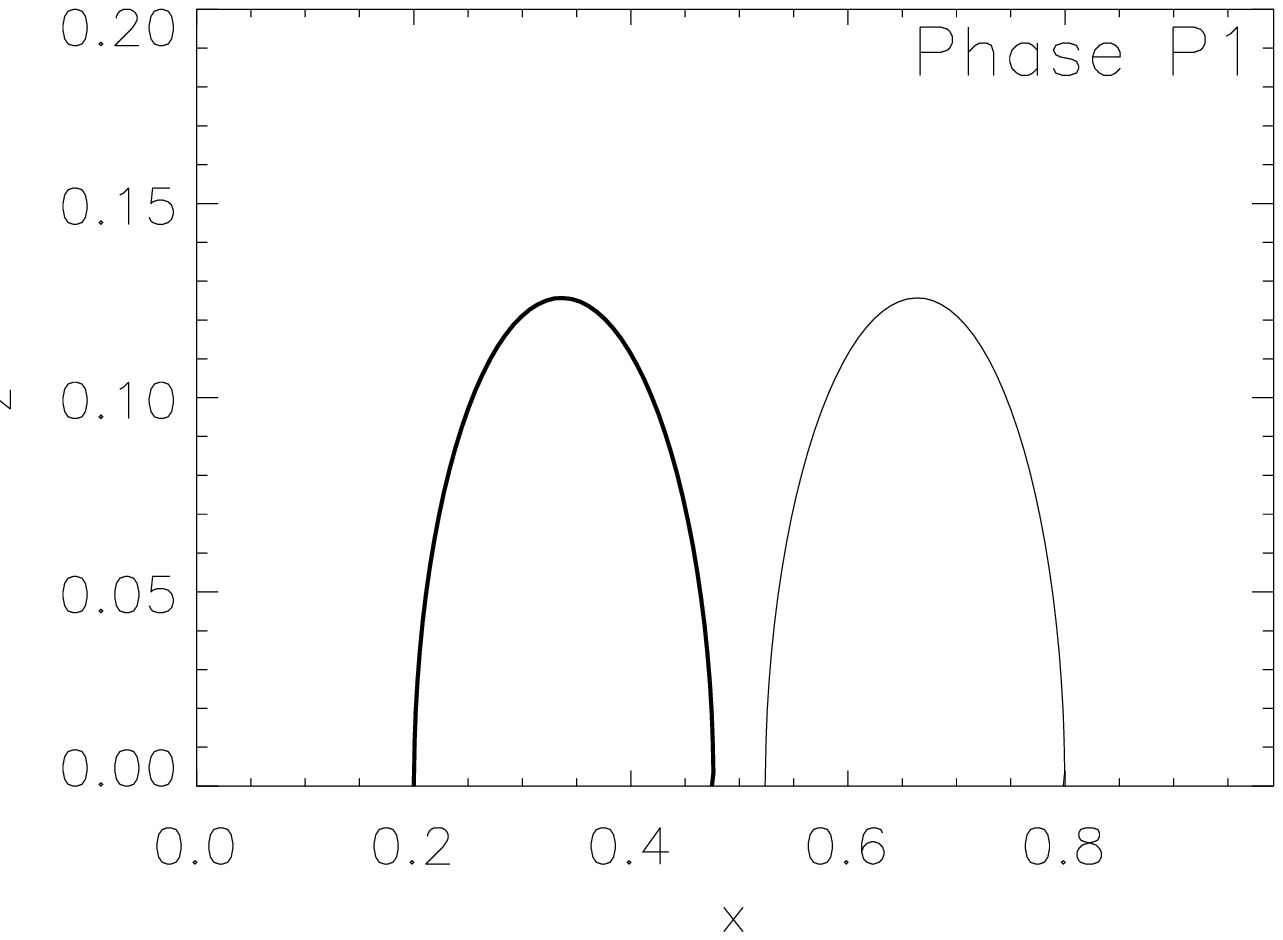}} \\
    \resizebox{.30\hsize}{.22\hsize}{%
        \includegraphics{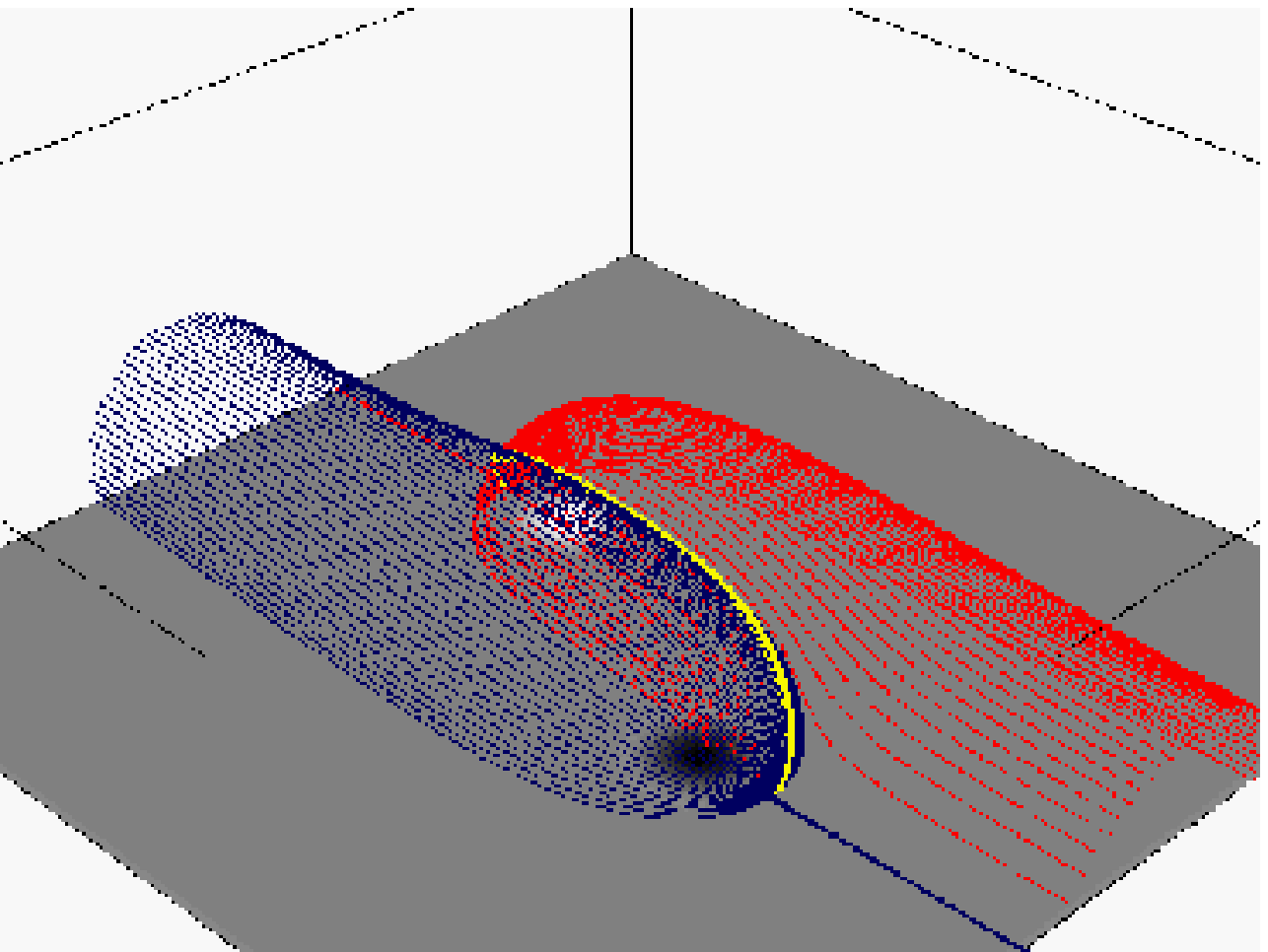}} &
      \resizebox{.30\hsize}{.22\hsize}{%
        \includegraphics{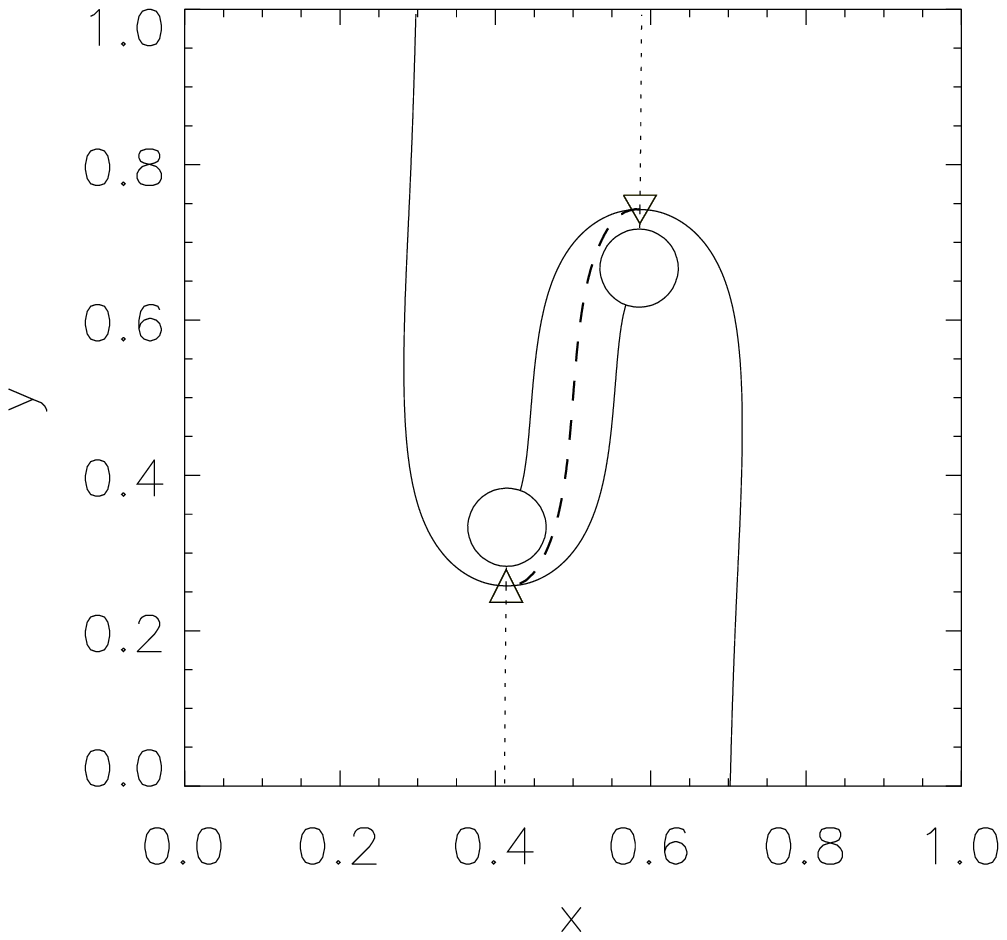}} &
      \resizebox{.30\hsize}{.22\hsize}{%
        \includegraphics{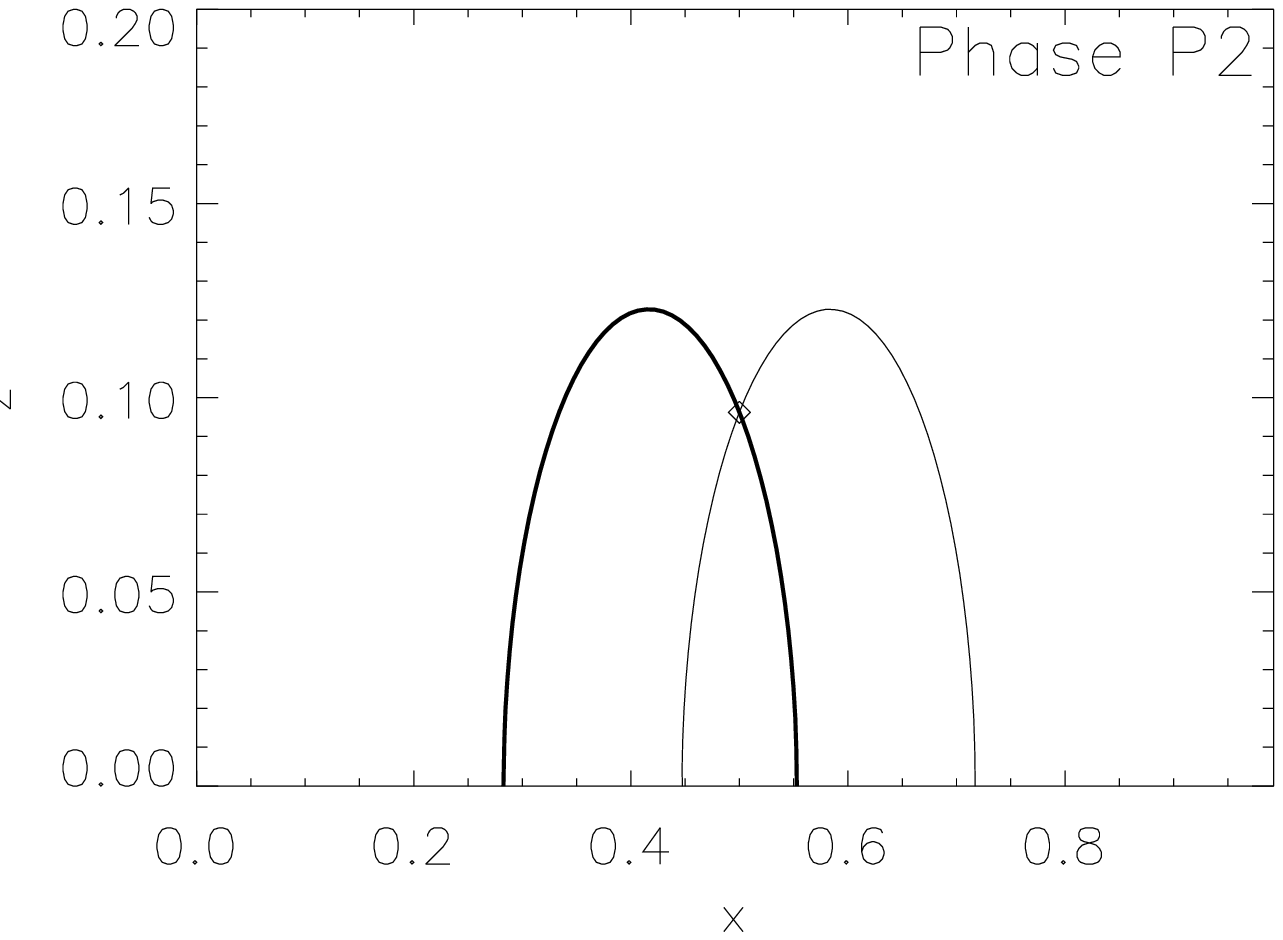}} \\
    \resizebox{.30\hsize}{.22\hsize}{%
        \includegraphics{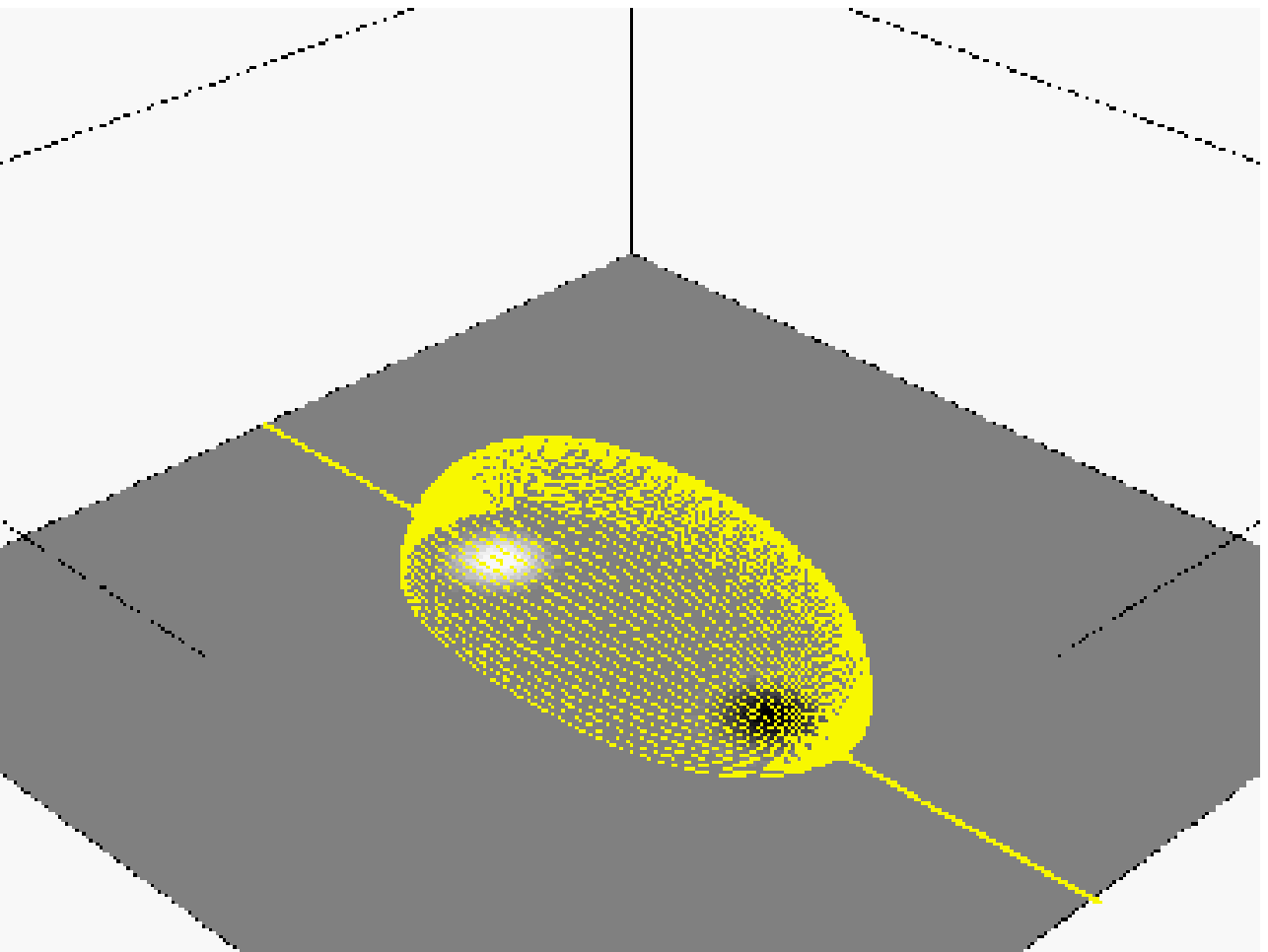}} &
      \resizebox{.30\hsize}{.22\hsize}{%
        \includegraphics{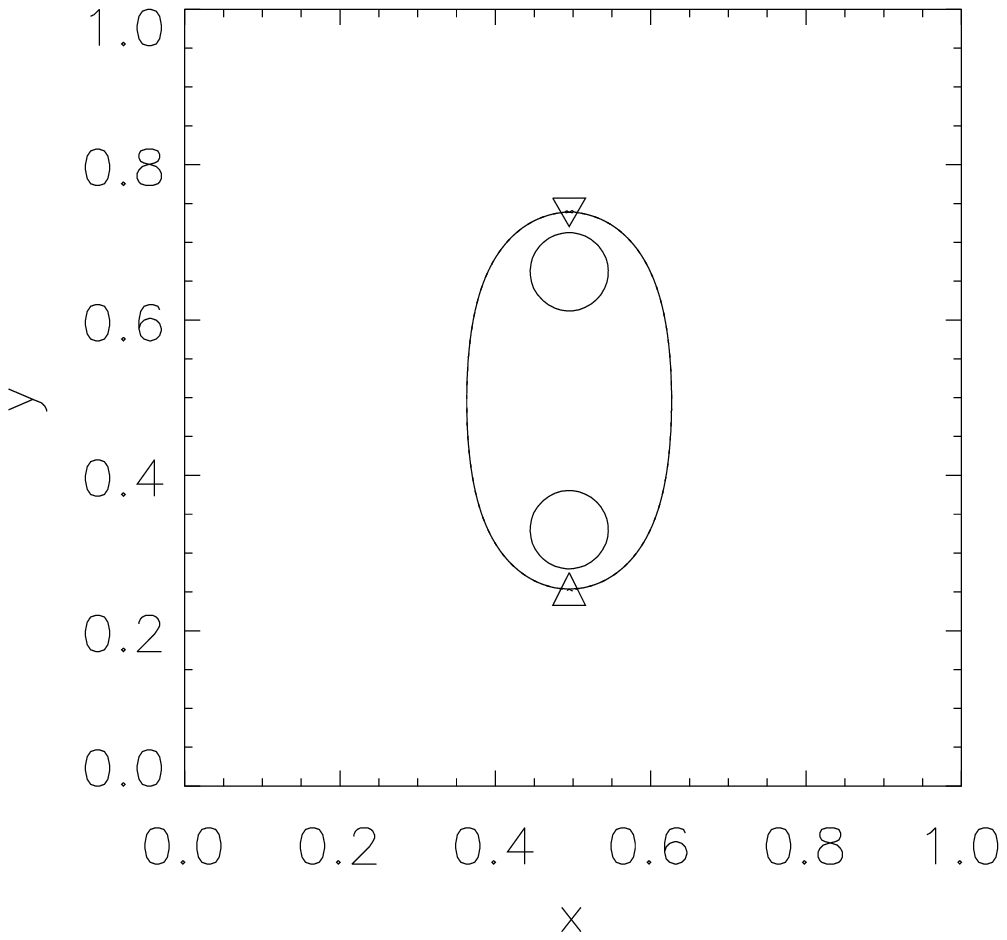}} &
      \resizebox{.30\hsize}{.22\hsize}{%
        \includegraphics{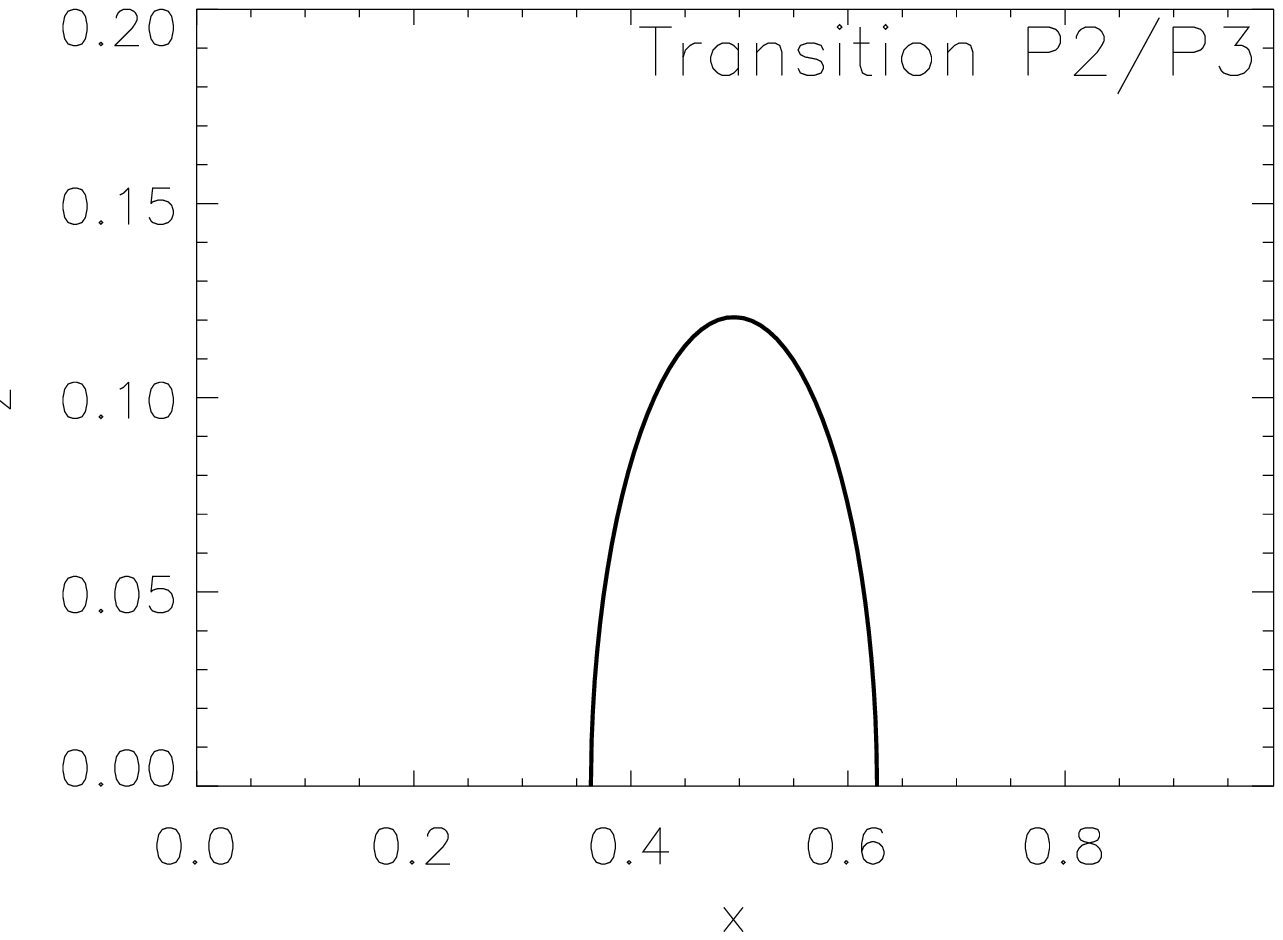}} \\
    \resizebox{.30\hsize}{.22\hsize}{%
        \includegraphics{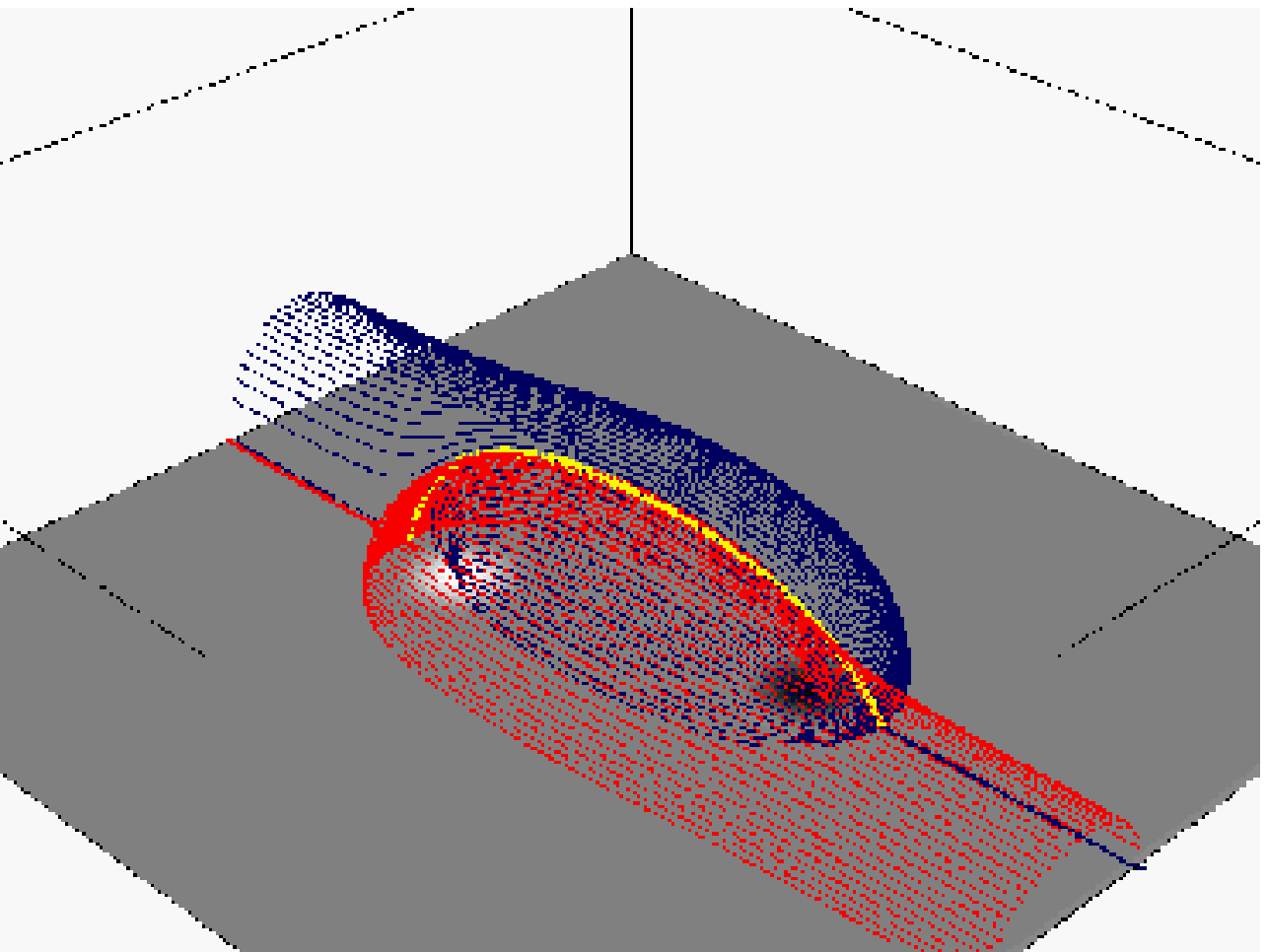}} &
      \resizebox{.30\hsize}{.22\hsize}{%
        \includegraphics{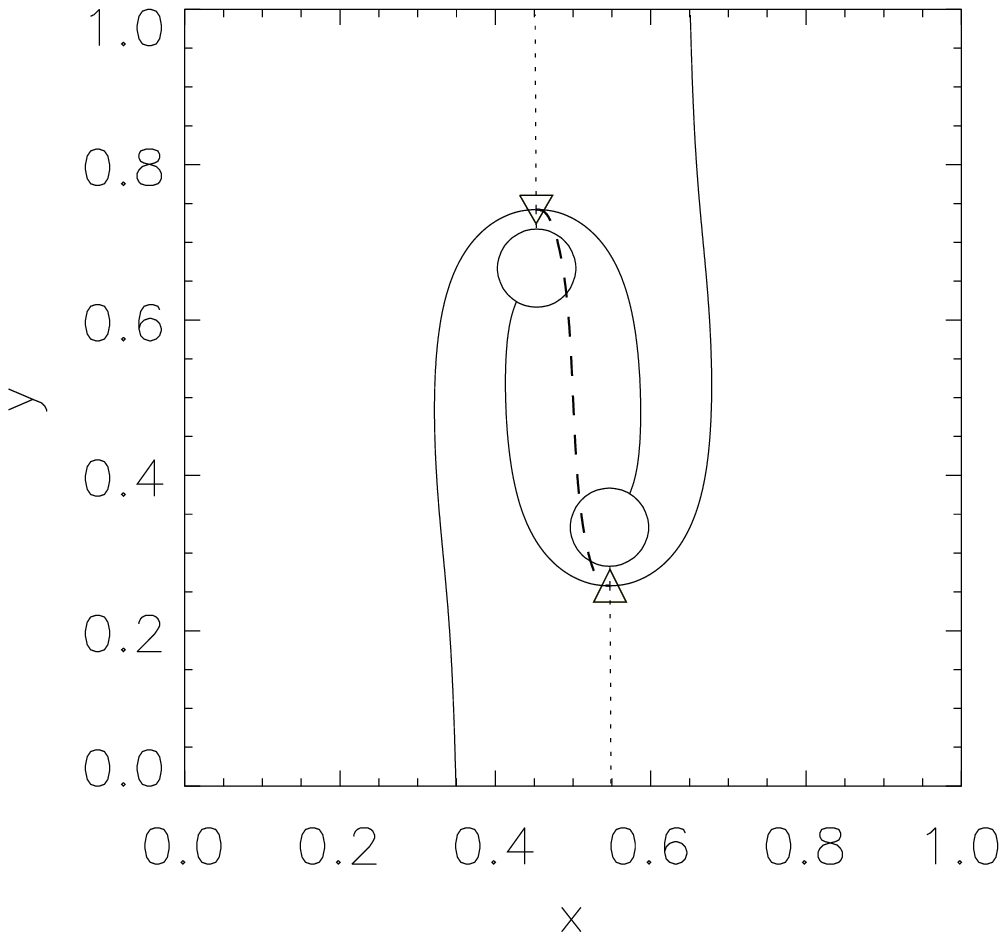}} &
      \resizebox{.30\hsize}{.22\hsize}{%
        \includegraphics{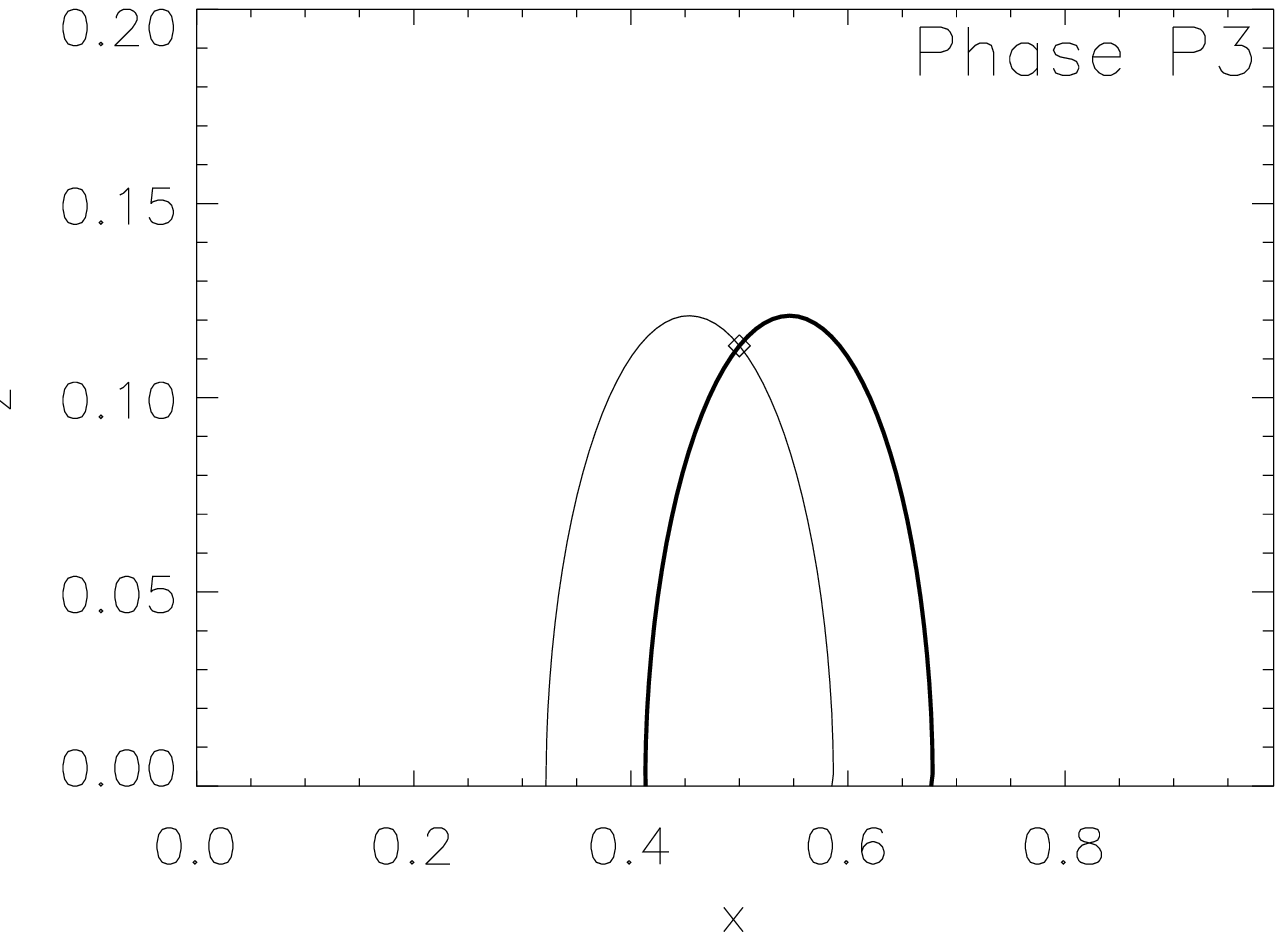}} \\
    \resizebox{.30\hsize}{.22\hsize}{%
        \includegraphics{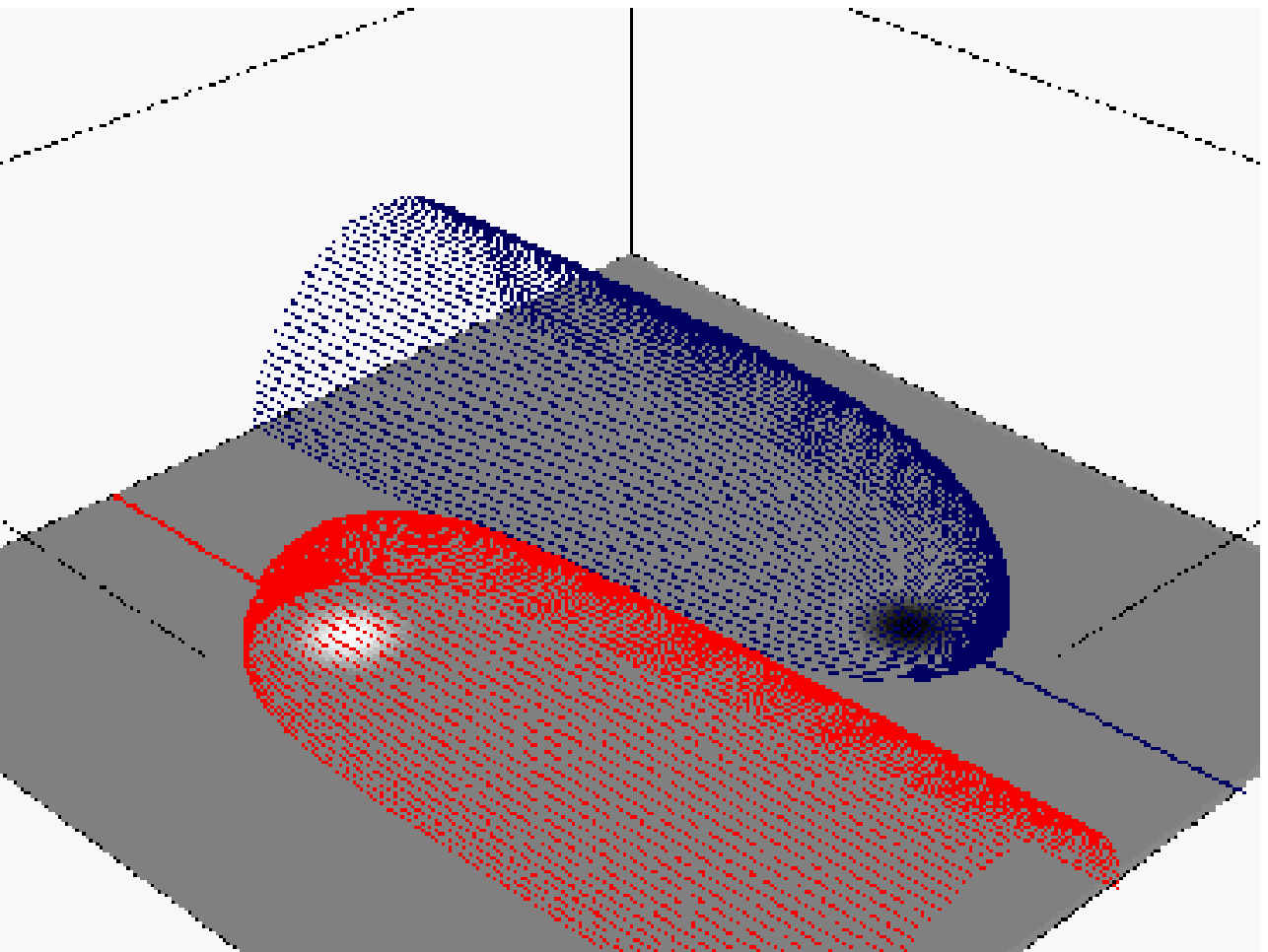}} &
      \resizebox{.30\hsize}{.22\hsize}{%
        \includegraphics{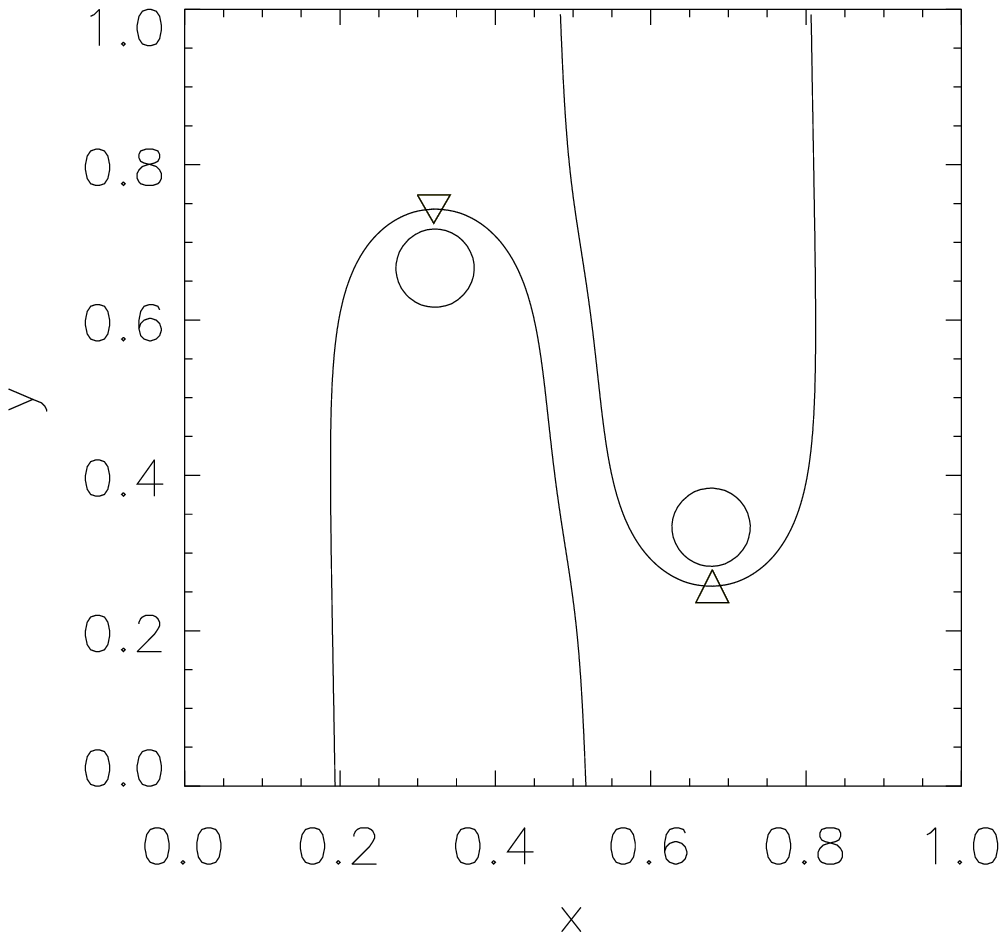}} &
      \resizebox{.30\hsize}{.22\hsize}{%
        \includegraphics{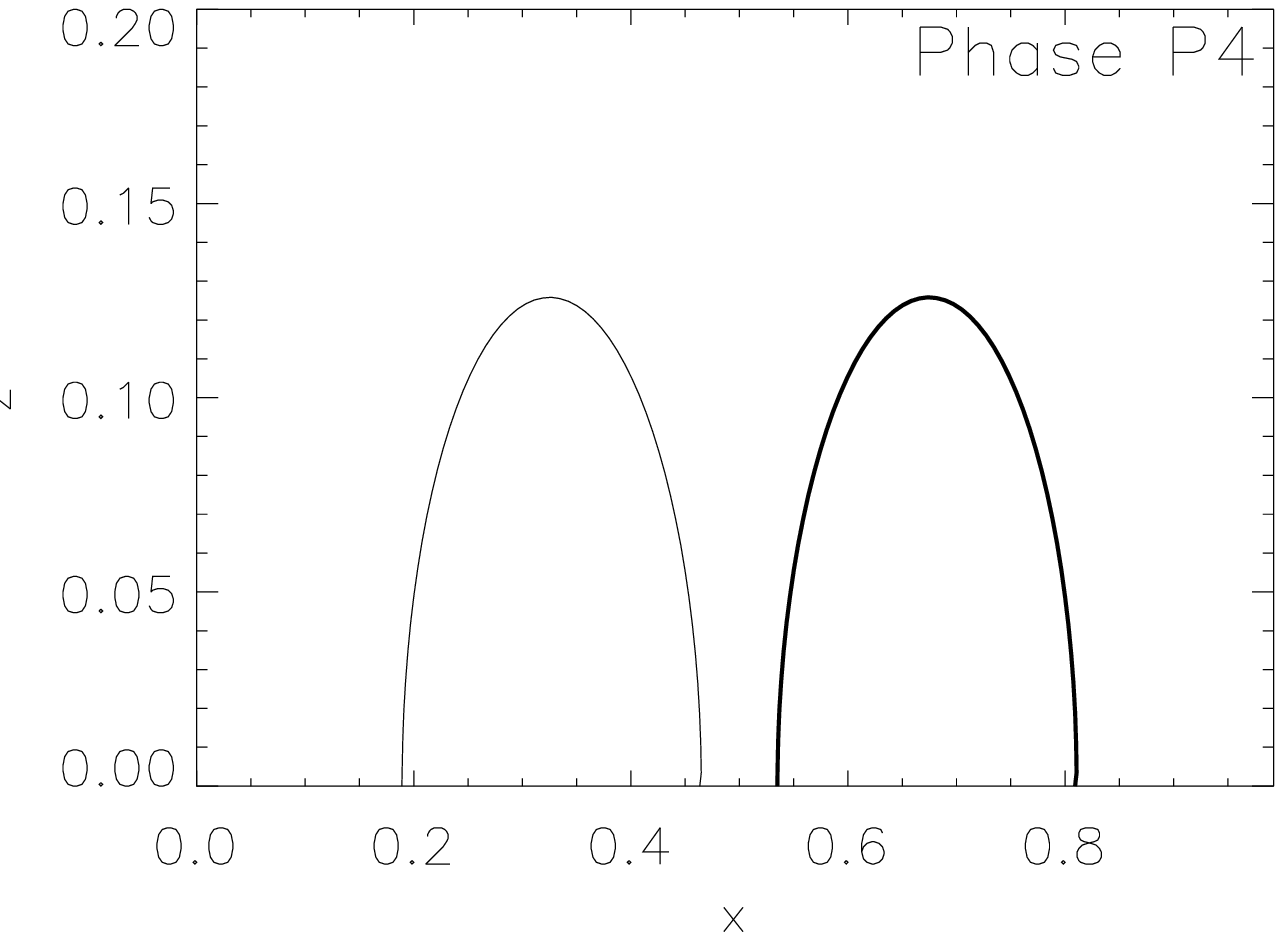}}
  \end{tabular}
  \caption{Snapshots of the skeletons for the potential model at $ t = 0.0,
    4.1, 8.14, 10.5, 19.9$ Alfv\'en times.  Column 1: 3D view, with positive separatrix surface
    (grey lines), negative separatrix surface (black lines) and separators
    (white lines).  Column 2: Footprint ($0 \leq x \leq 1, 0 \leq y \leq
    1$) at $z=0$, showing sources (hollow black circles), null points (positive -- $\bigtriangledown$ and negative -- $\bigtriangleup$) and
    intersections of separatrix surfaces with the photosphere (solid lines). The dashed lines show the separators above the source plane for comparison.
    Column 3: Cross-section ($0 \leq x \leq 1, 0 \leq z \leq 1/5 $) at
    $y=0.5$ showing the plane's intersection with the separatrix surfaces from the
    positive null (thin line) and the negative null (thick line)
    and with the separator ($\diamond$). 
  }
  \label{fig:pot-phases}
\end{figure}

\begin{figure}[ht]
  \resizebox{7cm}{!}{\includegraphics{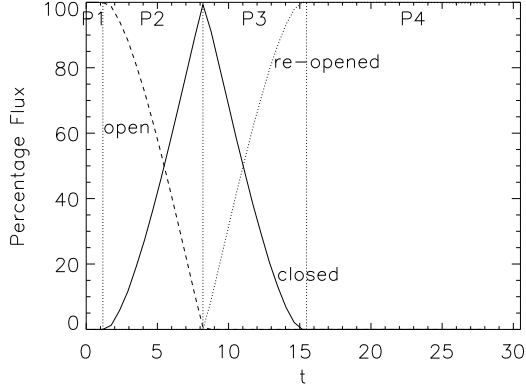}}
  \caption{
    Percentage of open (dashed), closed (solid) and reopened (dotted) flux 
    from the negative source versus the time ($t$) measured in Alfv\'en times 
    for the potential model (from Parnell \& Galsgaard 2004).  Vertical lines 
    represent the transition times between phases.}
  \label{fig:pot-open}
\end{figure}

The first phase has three flux domains: overlying, positive open and negative
open, where the positive source lies to the right of the negative source (figure~\ref{fig:pot-phases}, row 1).  There are no
separators and no reconnection occurs. Thus all the 
flux is open, as seen in figure~\ref{fig:pot-open}.  This phase lasts while the sources and their flux domains are pulled towards
each other, but ends as soon as the domains touch.
\clearpage
\clearpage
\begin{table}
  \caption{Characteristics of the potential phases}
  \longcaption{The components of each phase for the potential model, 
    namely, the number of separators ($X$), the multiplicity of each 
    source pair and the total number of flux domains (${\cal D}$).}
  \begin{tabular}{cc*4cc} \hline
    Phase & Seps.& \multicolumn{4}{c}{Source Pairs -- Flux Domain} & Total \\
    & ($X$)& over.  & +ve open & -ve open & closed & (${\cal D}$) \\ \hline
    P1 & 0        & 1  & 1  & 1 & 0  & 3     \\
    P2 & 1        & 1  & 1  & 1 & 1 & 4     \\
    P3 & 1        & 1  & 1  & 1  & 1  & 4    \\
    P4 & 0        & 1  & 1  & 1  & 0  & 3     \\ \hline
  \end{tabular}
  \label{tab:phases-pot} 
\end{table}
\subsubsection*{Phase P2 ($1.47<t<8.14$): Closing}
A global separator bifurcation occurs in which a separator rises 
from the source plane. This separator runs the ridge of a new domain that is 
made up of new reconnected closed flux. (figure~\ref{fig:pot-phases}, row 2).  
Reconnection creates closed flux and destroys open flux throughout this phase (figure~\ref{fig:pot-open}).
 The sources continue to move towards one another during this stage, until they reach
their point of closest approach and are aligned with the overlying field. 
At this point the two sources are completely connected, the closed flux 
reaches its maximum (figure~\ref{fig:pot-open}) and a new phase must start.
\subsubsection*{Phase P3 ($8.14<t<15.49$): Reopening}
As the separatrix surfaces of both nulls fully coincide, a 
\emph{global separatrix bifurcation} occurs in which an infinite number of 
separators completely encompasses the closed flux domain isolating it. This is 
a new type of bifurcation (figure~\ref{fig:pot-phases}, row 3). As soon as the 
sources move out of alignment with the overlying field the flux reopens 
through reconnection and a new single separator state is formed. The continued advection of the sources creates a 
positive open domain on the left and a negative open
domain on the right with a closed domain in between 
(figure~\ref{fig:pot-phases}, row 4).  These domains are all simply connected and surrounded by 
overlying field.  A separator runs the ridge of the closed domain, as in 
phase P2. It descends towards the base as closed flux (with overlying flux) 
is converted back into positive and negative open flux (see
 figure~\ref{fig:pot-open}).
This is the reverse of phase P2 when the closed
flux was formed. This
phase lasts until the separator has reached the base
\subsubsection*{Phase P4 ($15.49<t<26.54$): Final Open}
In this phase (figure~\ref{fig:pot-phases}, row 5) a global separator bifurcation destroys the
separator when it reaches the source plane, leaving just three simply connected flux domains: overlying,
positive open and negative open.  These flux domains are equivalent to those in phase P1, except that, as the sources have now passed each other, they are now
moving apart instead of together and the negative source now lies to the
right of the positive one. There is no reconnection in this phase (see figure~\ref{fig:pot-open}).
\subsection{Evolution of MHD skeleton} \label{sub:evol-mhd}

The dynamic MHD experiment is executed using the method described in
\S2b.  From this, the skeleton for each time frame is
deduced.  As in the potential model, we visualise the skeletons in 3D and also
in 2D using both the footprint at $z=0$ and a cross-section at $y=0.5$ 
(see figure~\ref{fig:mhd-phases}). A filled contour plot of current intensity 
is added to the cross-section.

As before, we group contiguous time
frames into phases of the same topology.  These phases are: D1 - initial open,
D2 - double-separator hybrid, D3 - single-separator closing,
D4 - quintuple-separator hybrid, D5 - triple-separator hybrid and D6 - single-separator reopening (table~\ref{tab:phases-mhd}).  We use a naming convention based on the number 
of separators and the 
reconnection process occurring, e.g. closing, hybrid (closing and reopening), reopening (see Parnell et al. 2007 for physical explanation).  The phases are described in turn. 
\subsubsection*{Phase D1 ($0.0<t<5.49$): Initial Open}

Since the magnetic field is assumed to be potential at $t=0$, phase D1 (figure~\ref{fig:mhd-phases}, row 1) is
topologically equivalent to phase P1.  Hence, there are no separators and three simply connected
flux domains, namely: overlying, positive open and negative open. The positive source (and its corresponding flux domain) lies to the right of
the negative source (and its flux domain).

As the positive open and negative open flux domains are advected towards one 
another the two flux domains push up against each other, save for a small gap below (figure~\ref{fig:mhd-phases}, 
row 1, col 2).  At the time this phase ends, about half of the closed flux in
the potential experiment has been reconnected (figure~\ref{fig:pot-open}).  Thus, since, up to this point, there has been no reconnection in the dynamic 
experiment, it is natural that a thin narrow current sheet has developed by the
end of this phase.

\subsubsection*{Phase D2 ($5.49<t<9.68$): Double-Separator Hybrid}

\begin{figure}[ht]  
  \begin{tabular}{ccc}
    \resizebox{.30\hsize}{.22\hsize}{%
        \includegraphics{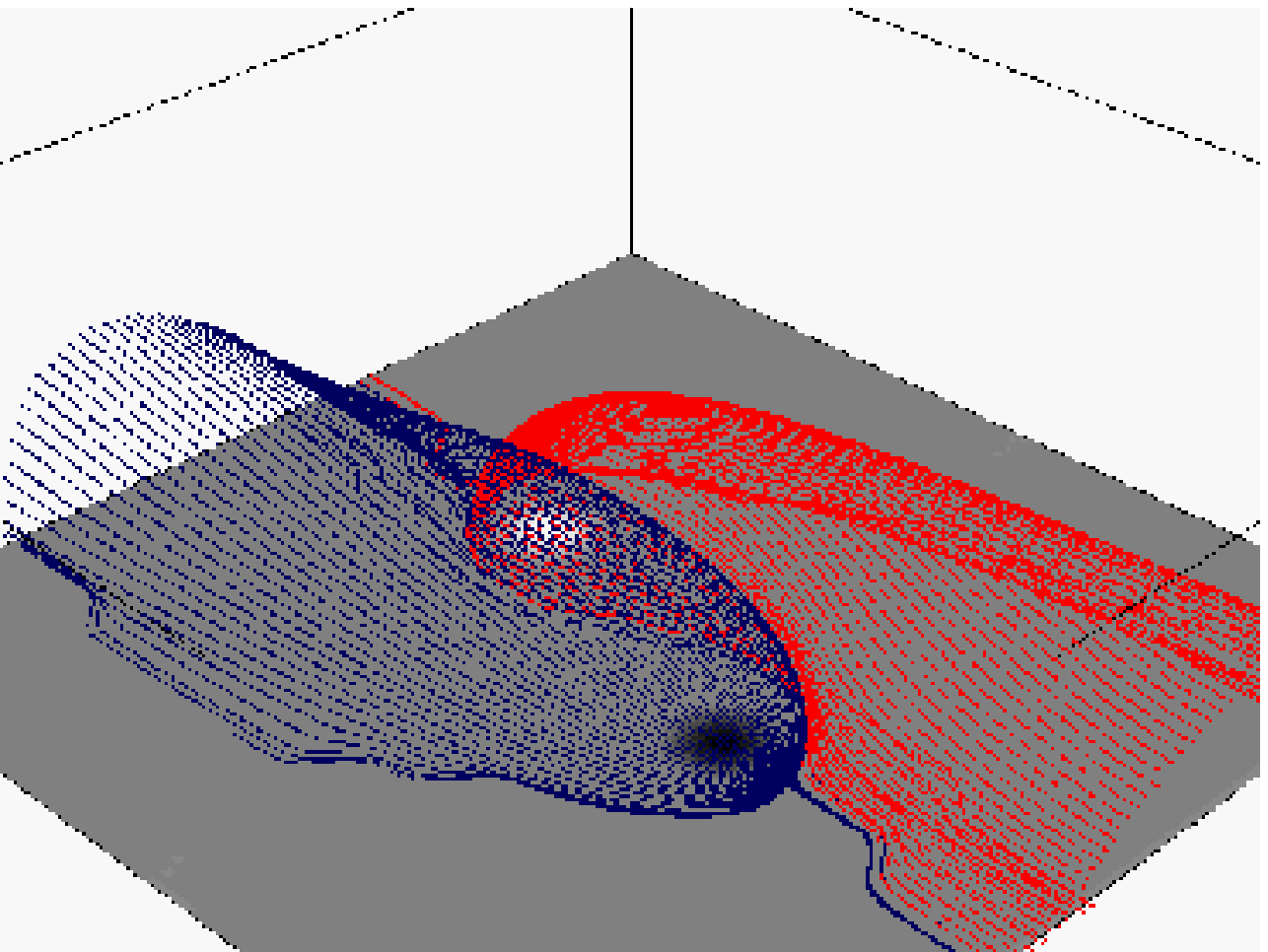}} &
      \resizebox{.30\hsize}{.22\hsize}{%
        \includegraphics{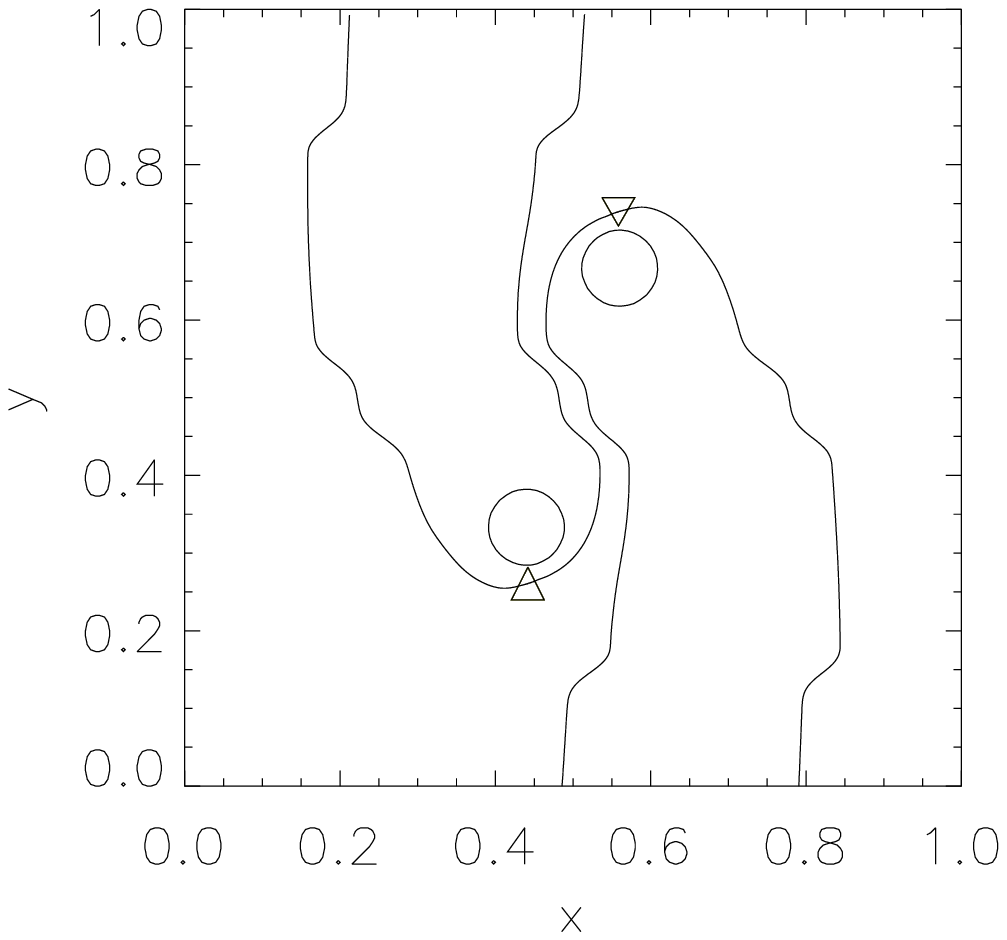}} &
      \resizebox{.30\hsize}{.22\hsize}{%
        \includegraphics{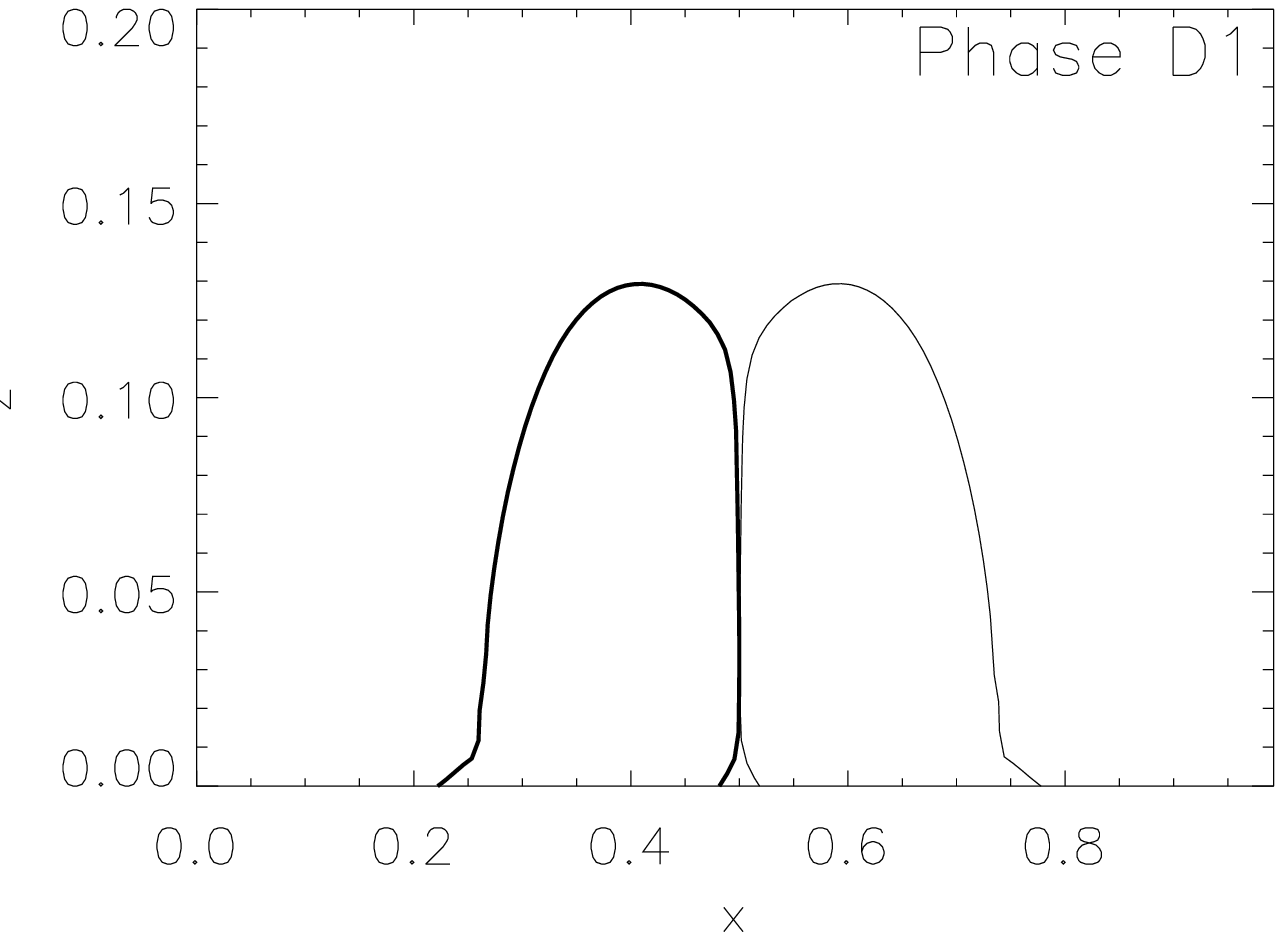}} \\
    \resizebox{.30\hsize}{.22\hsize}{%
        \includegraphics{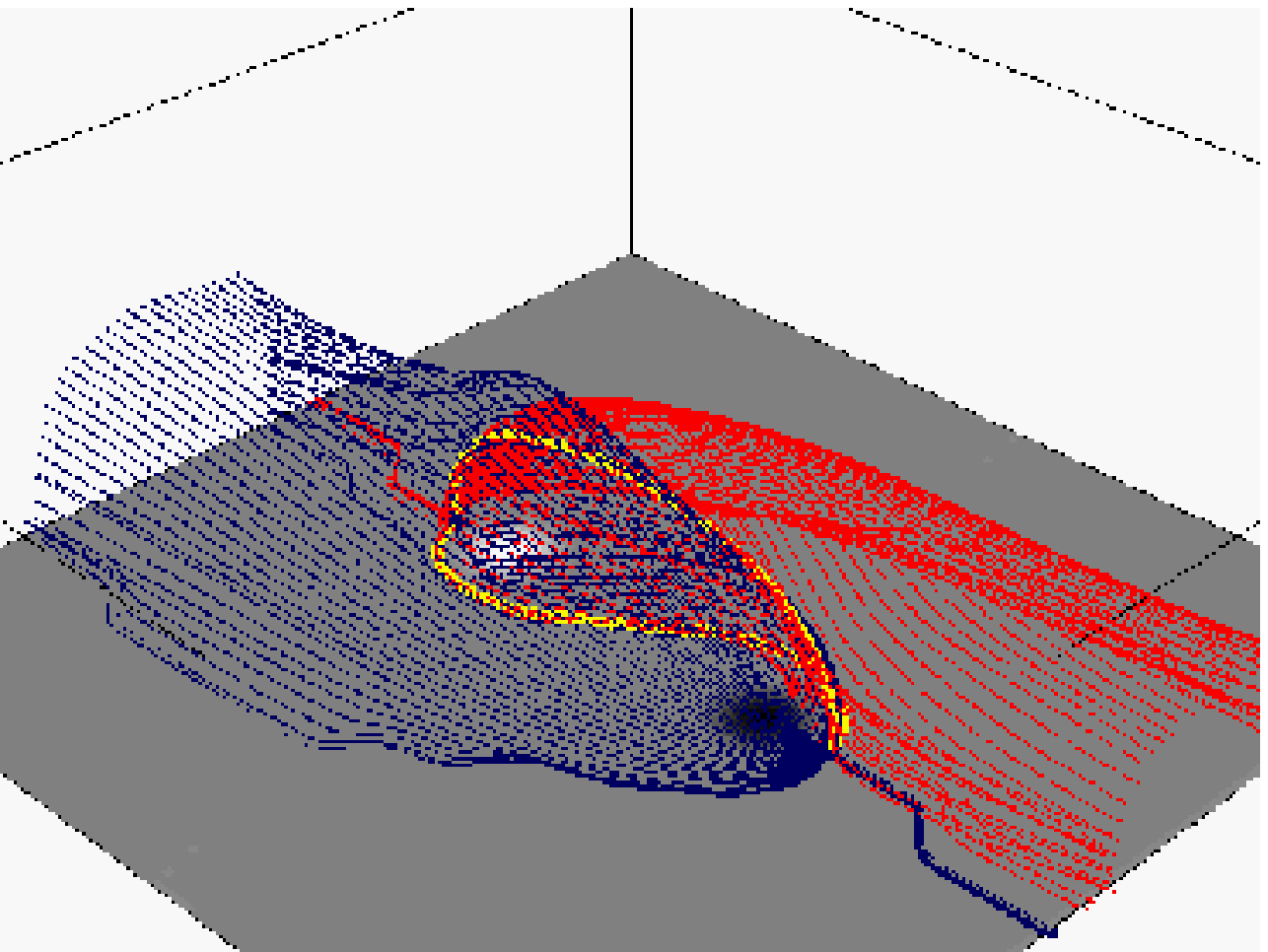}} &
      \resizebox{.30\hsize}{.22\hsize}{%
        \includegraphics{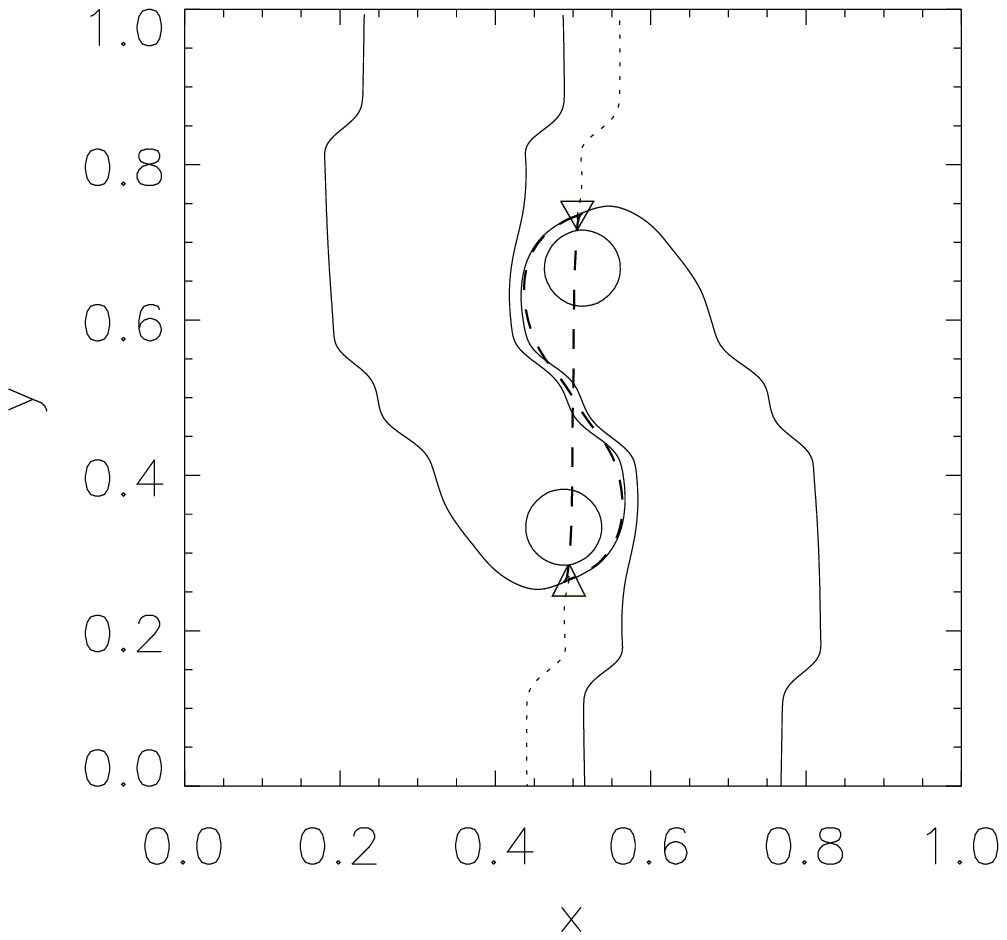}} &
      \resizebox{.30\hsize}{.22\hsize}{%
        \includegraphics{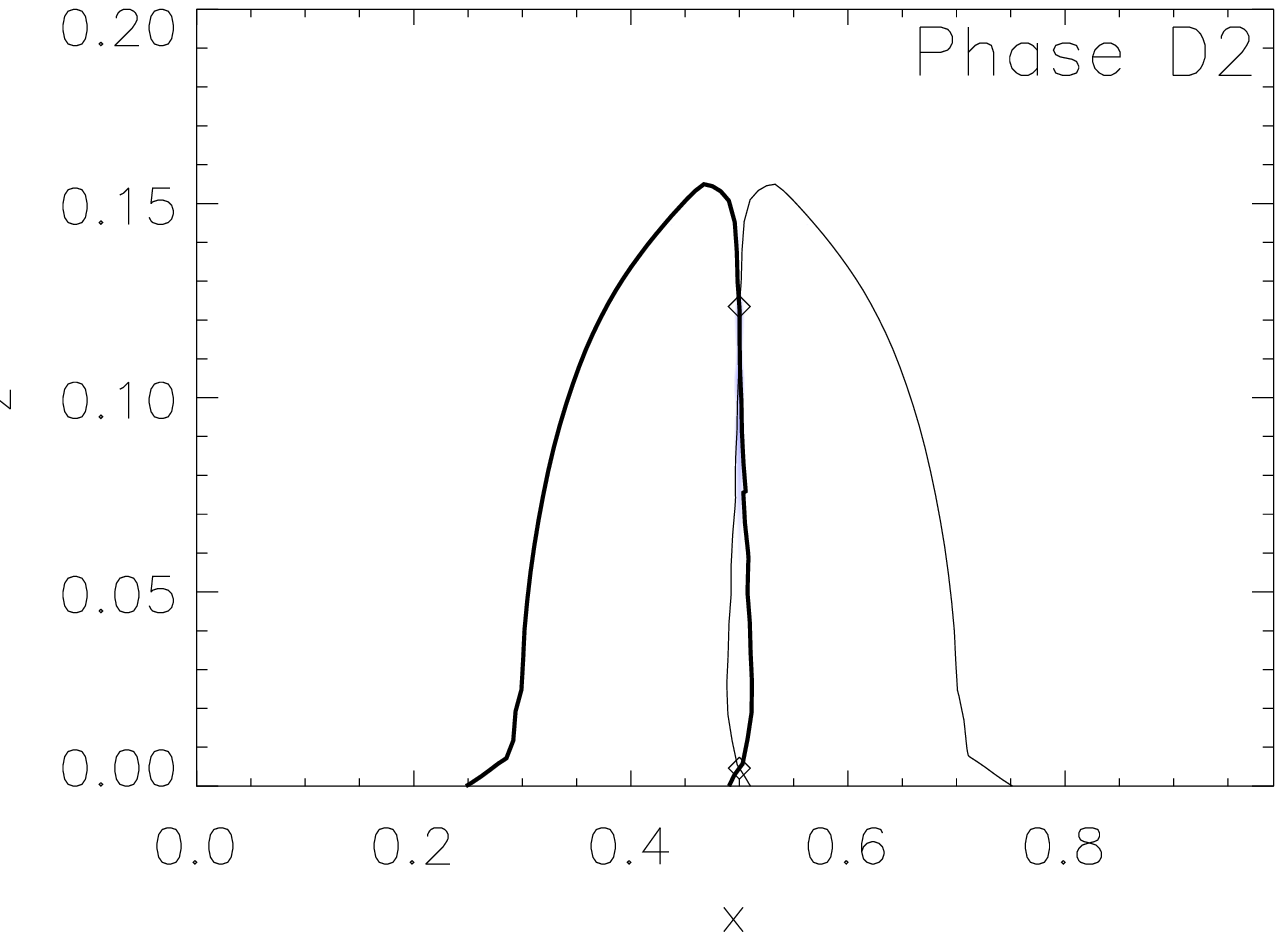}} \\
    \resizebox{.30\hsize}{.22\hsize}{%
        \includegraphics{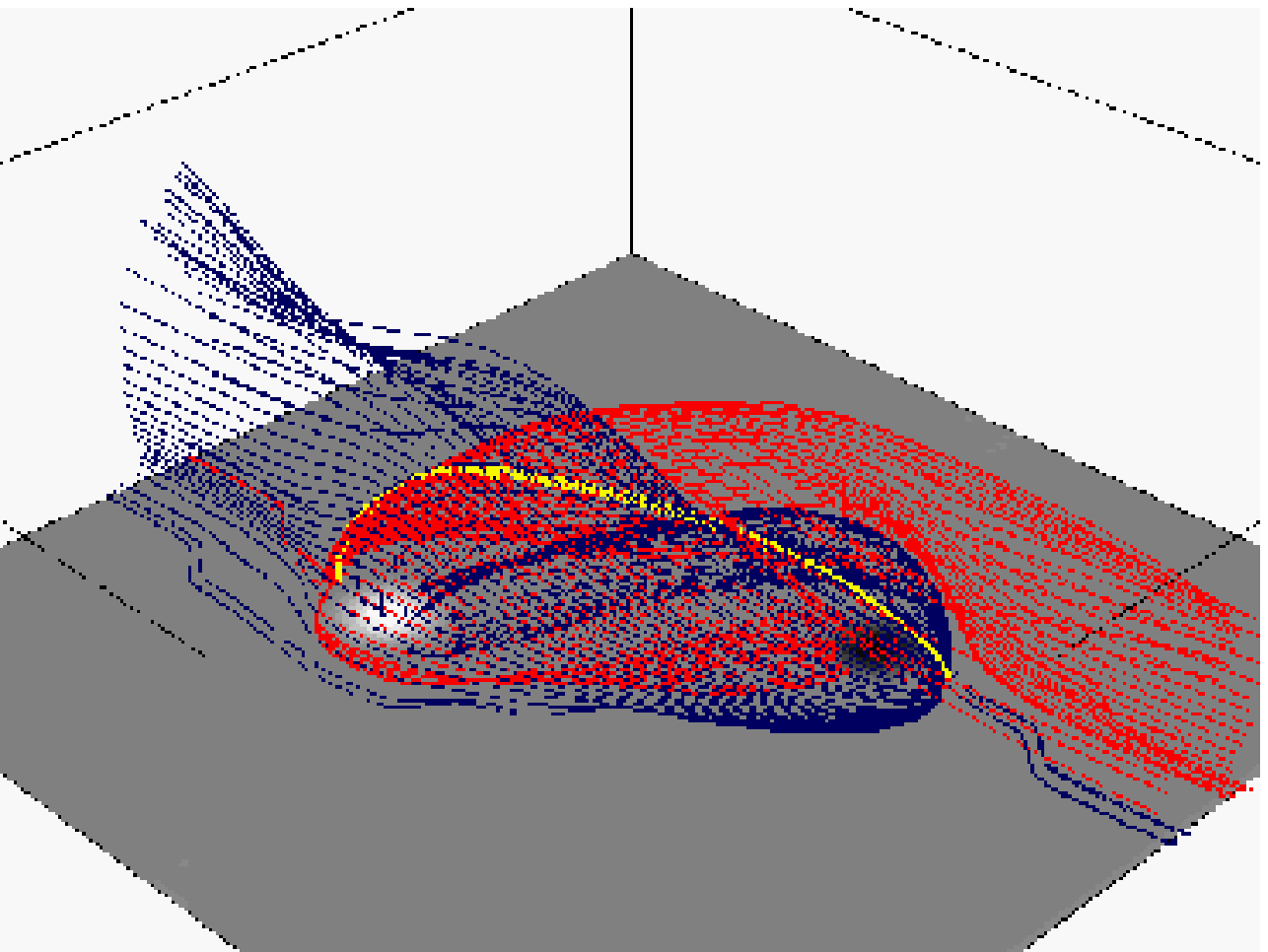}} &
      \resizebox{.30\hsize}{.22\hsize}{%
        \includegraphics{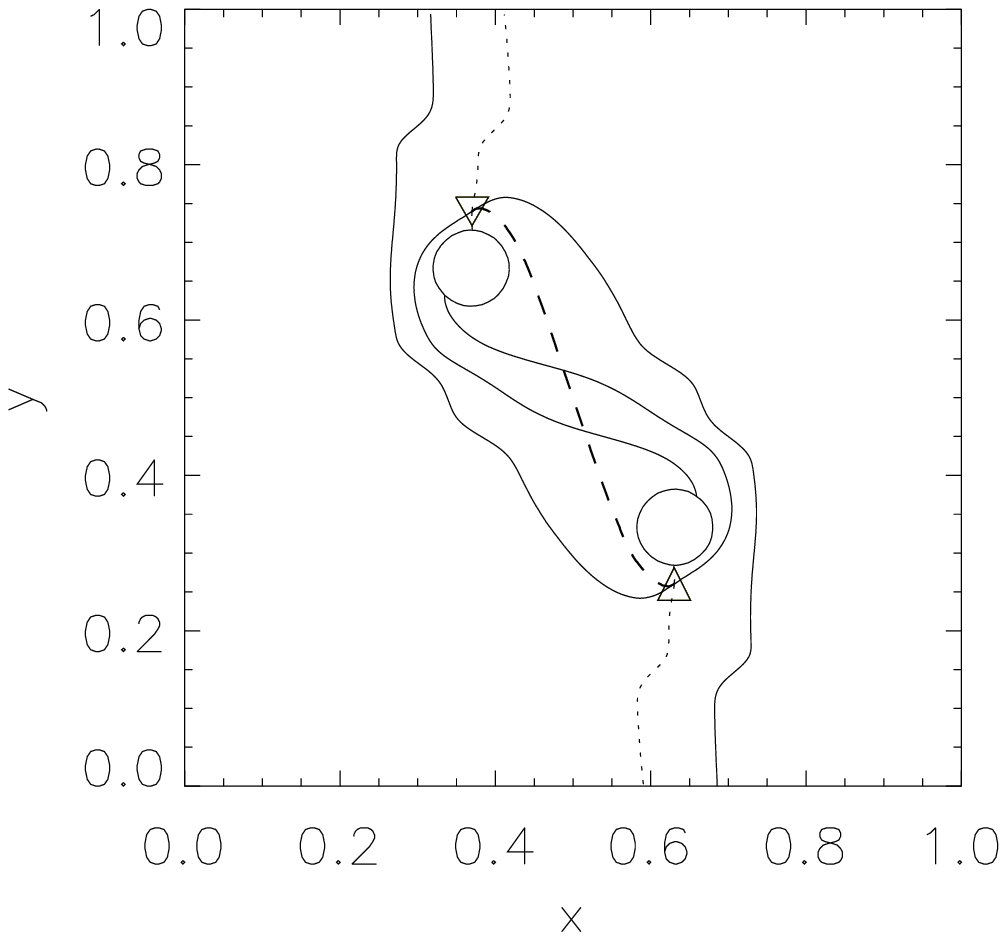}} &
      \resizebox{.30\hsize}{.22\hsize}{%
        \includegraphics{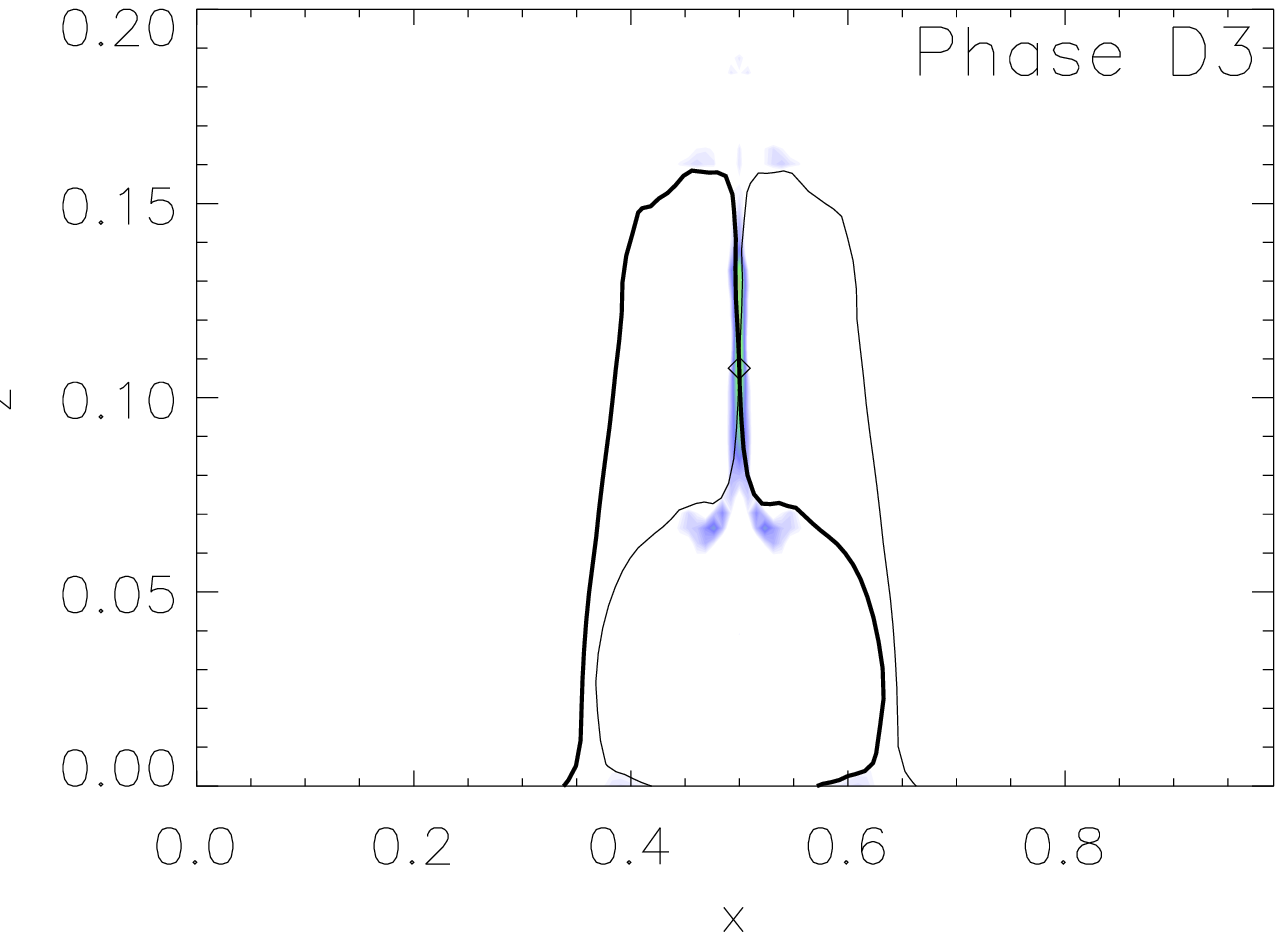}} \\
    \resizebox{.30\hsize}{.22\hsize}{%
        \includegraphics{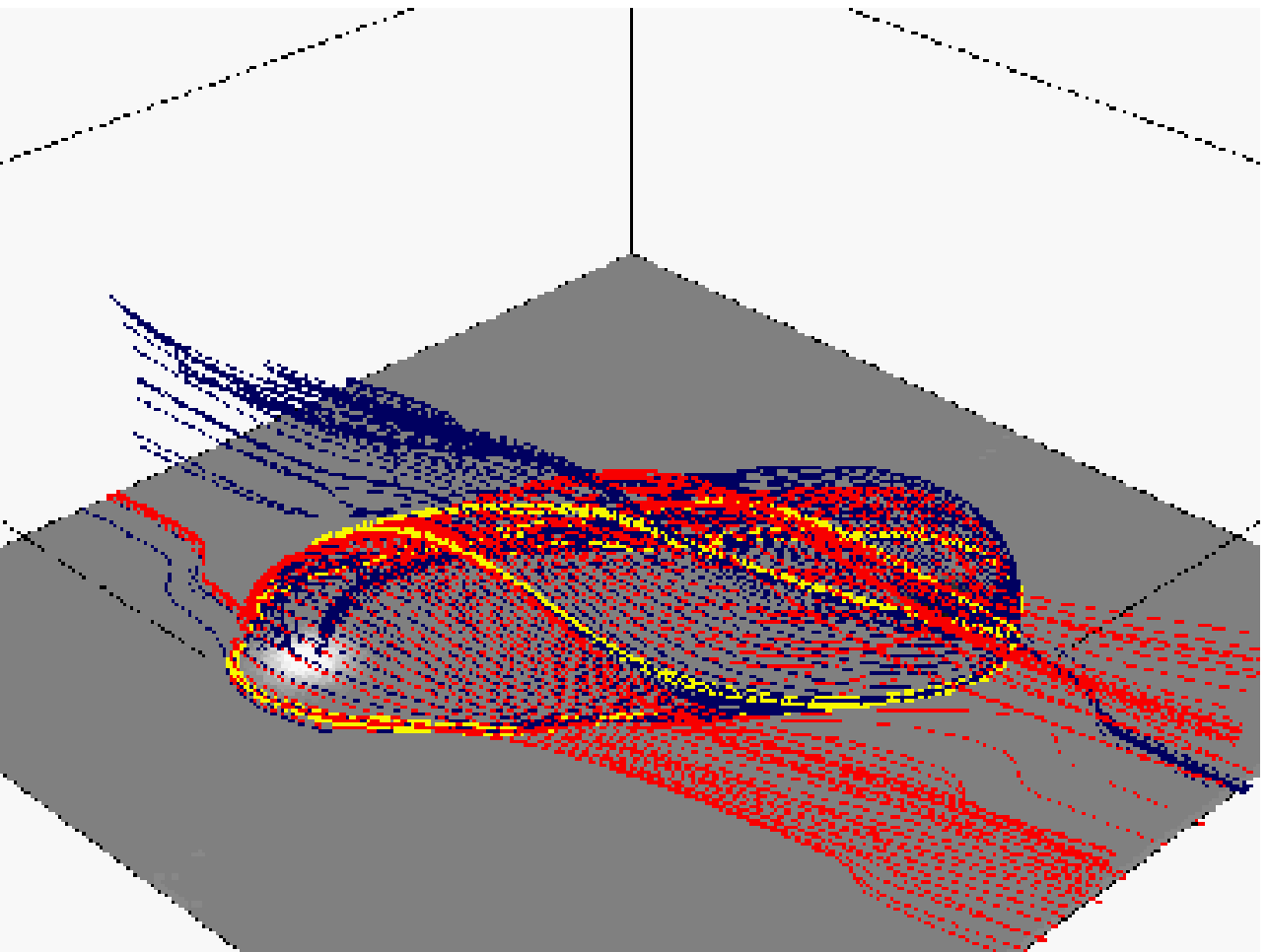}} &
      \resizebox{.30\hsize}{.22\hsize}{%
        \includegraphics{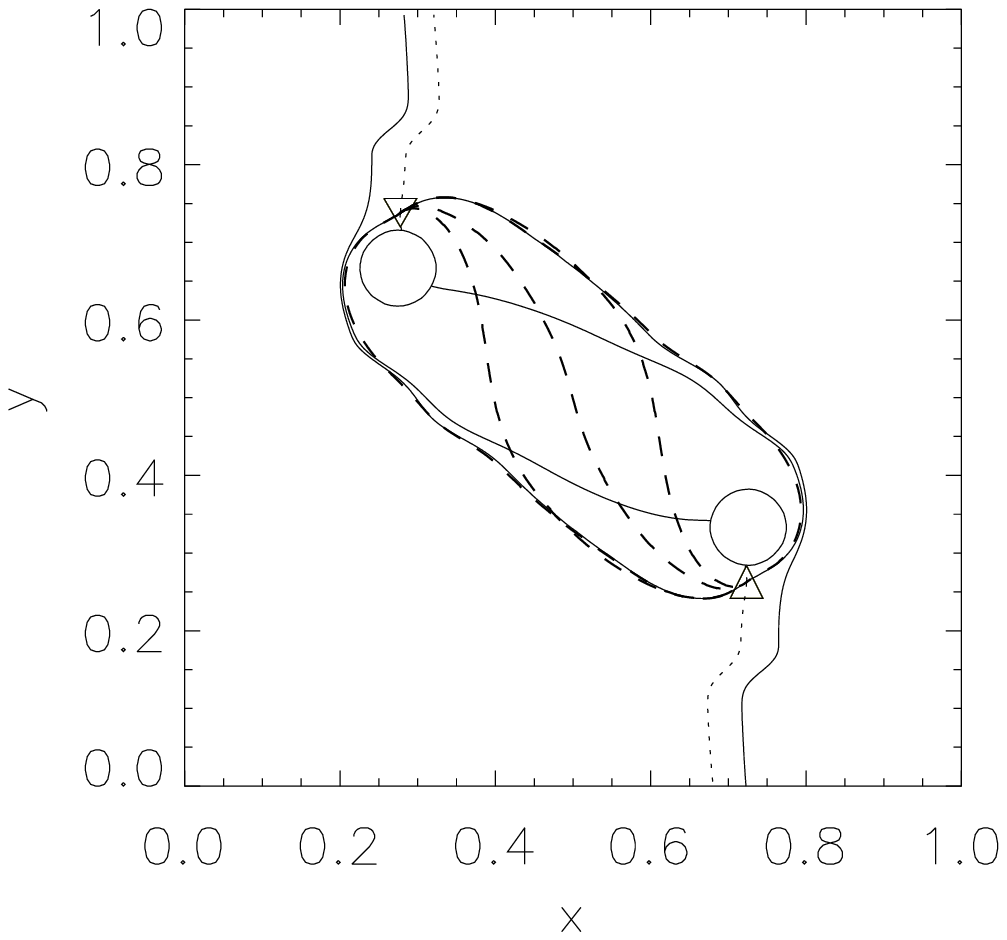}} &
      \resizebox{.30\hsize}{.22\hsize}{%
        \includegraphics{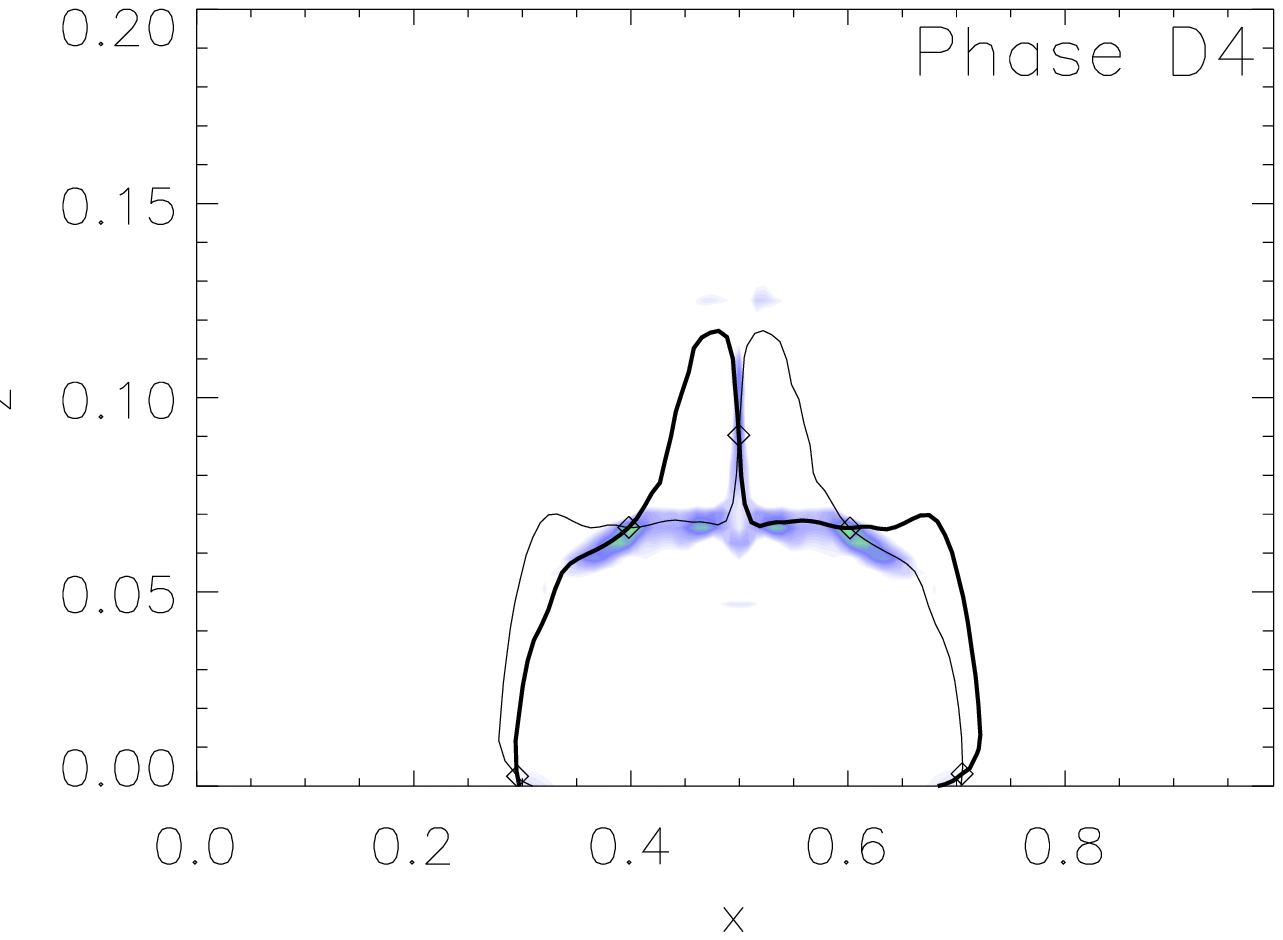}} \\
    \resizebox{.30\hsize}{.22\hsize}{%
        \includegraphics{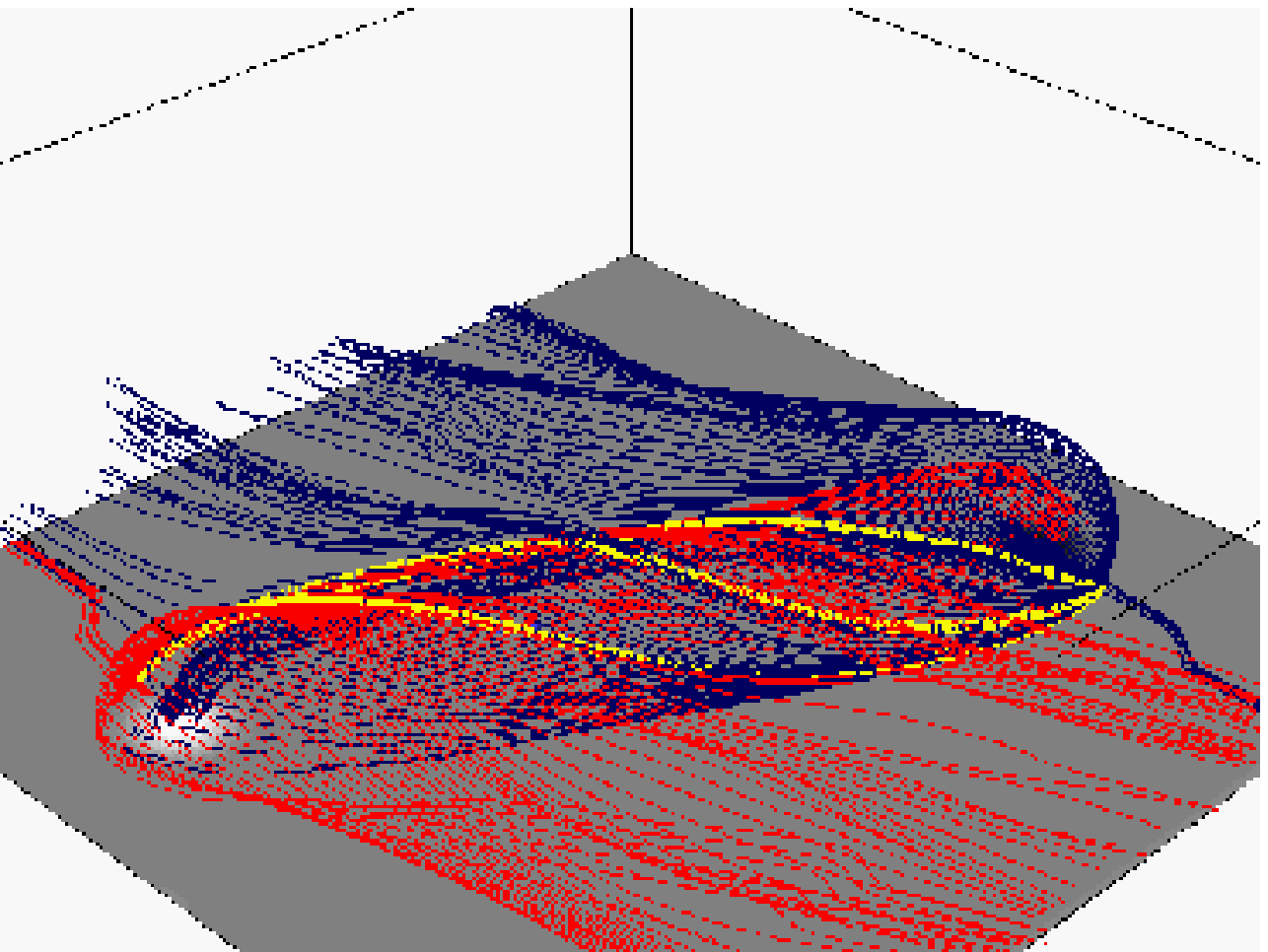}} &
      \resizebox{.30\hsize}{.22\hsize}{%
        \includegraphics{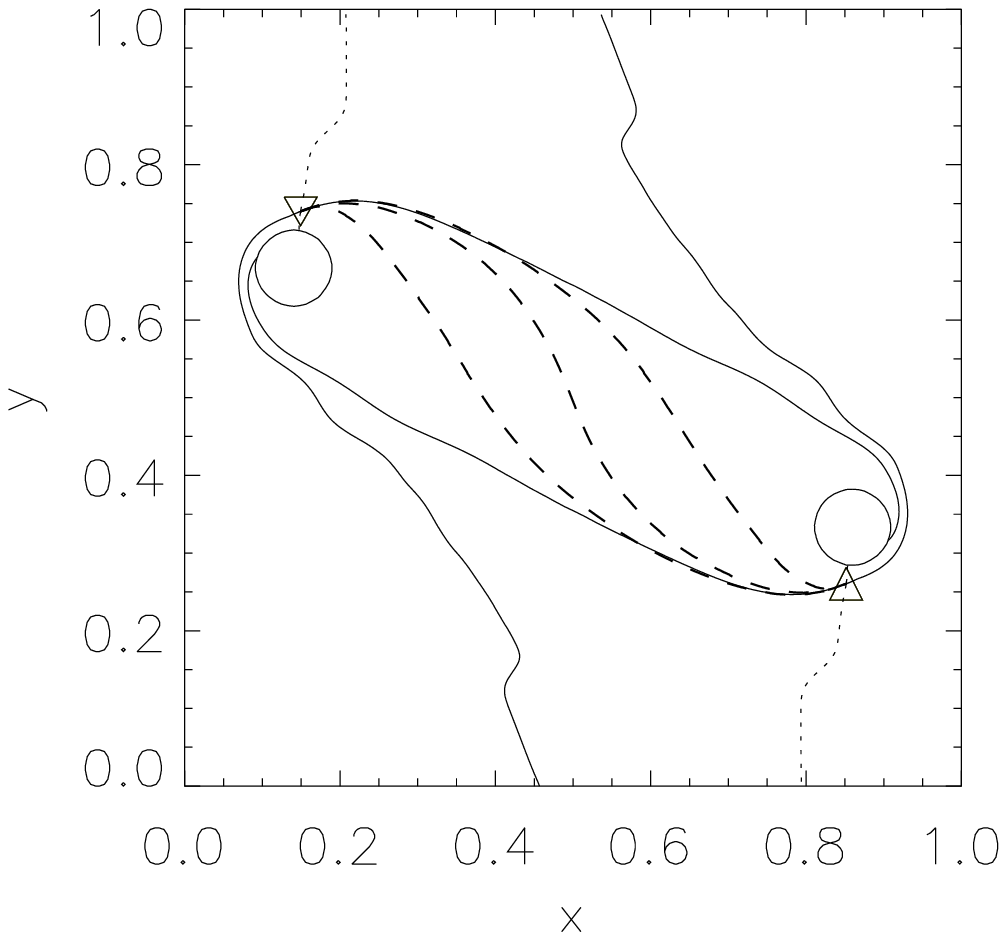}} &
      \resizebox{.30\hsize}{.22\hsize}{%
        \includegraphics{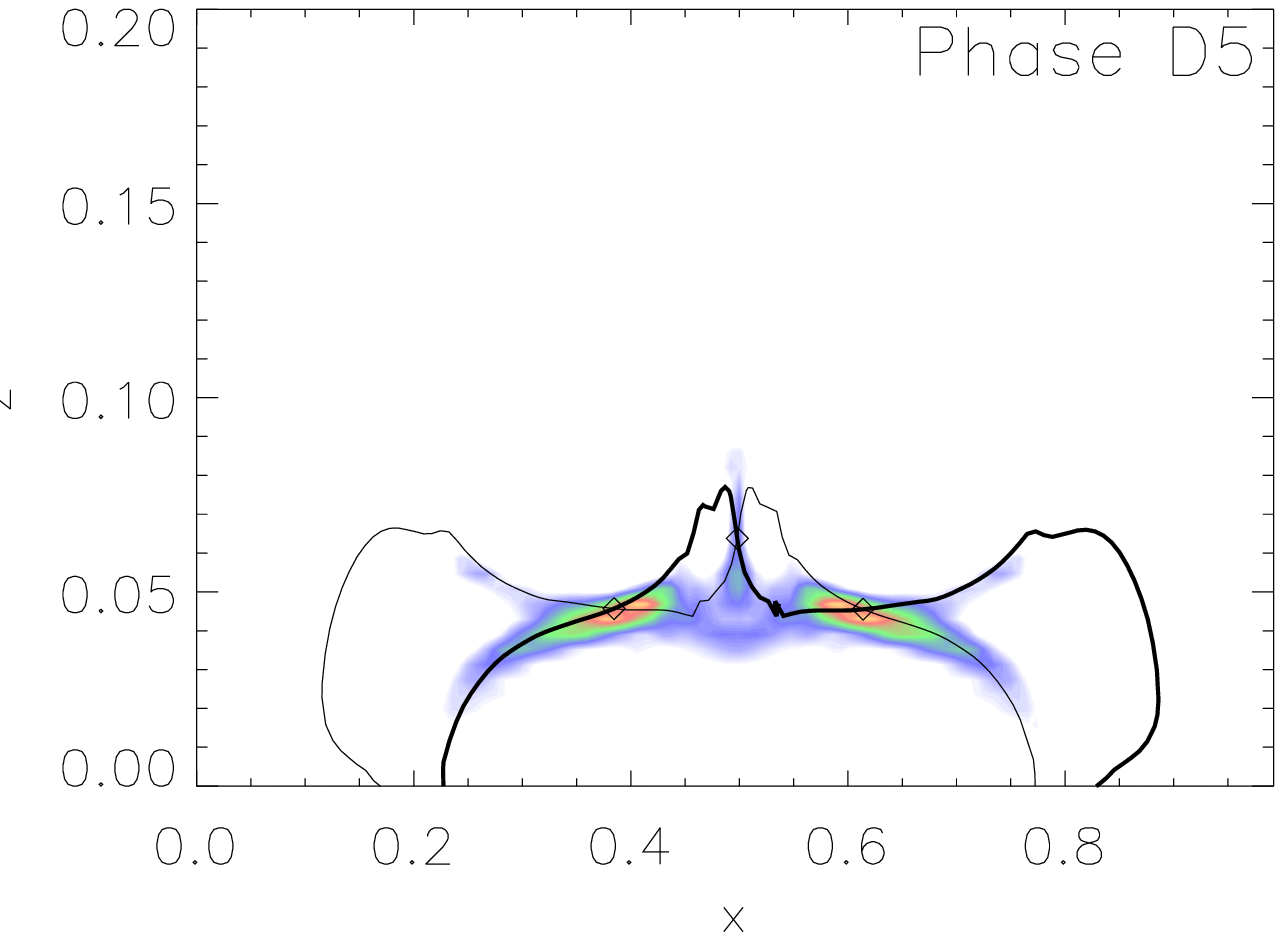}} \\
    \resizebox{.30\hsize}{.22\hsize}{%
        \includegraphics{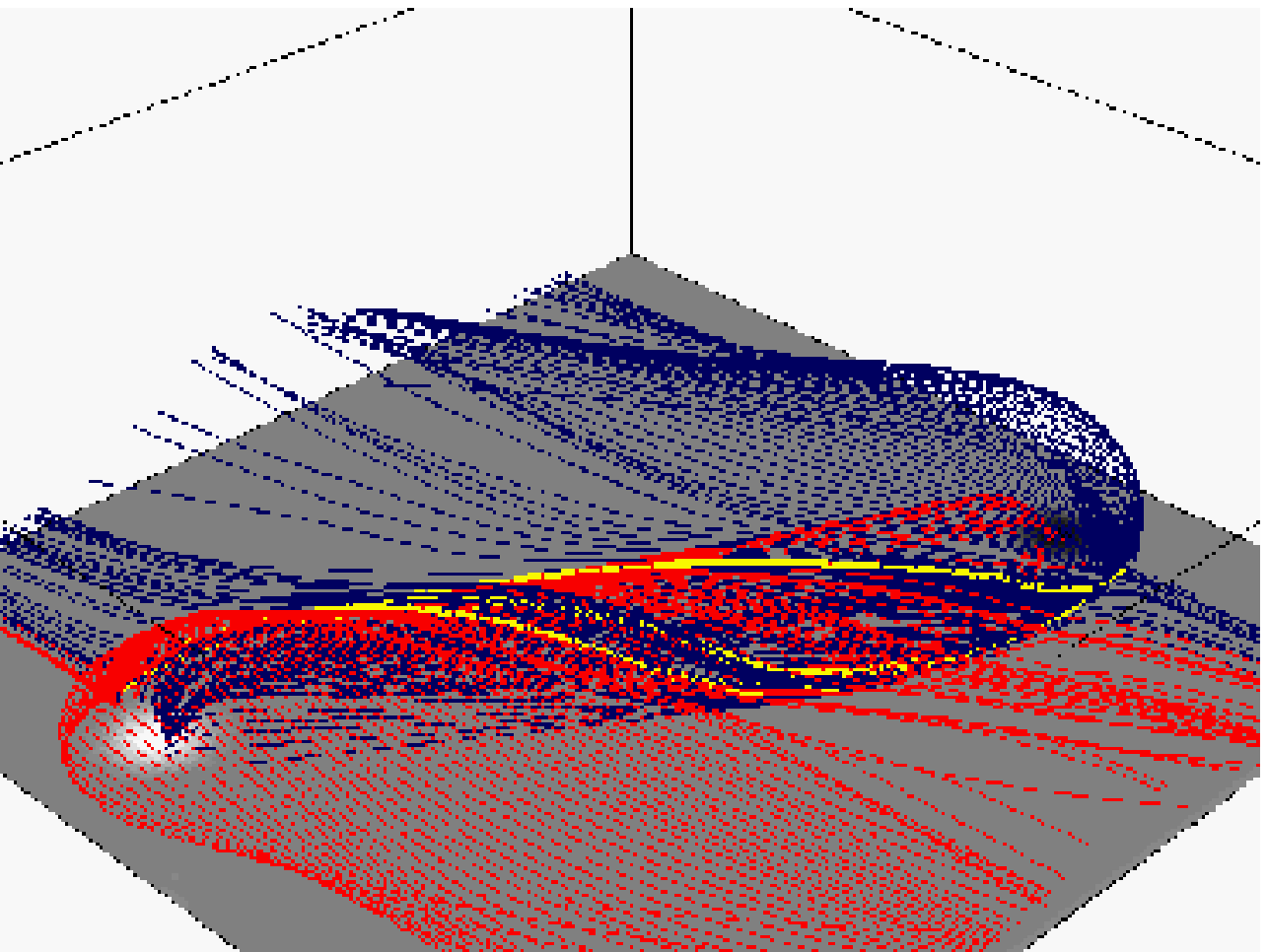}} &
      \resizebox{.30\hsize}{.22\hsize}{%
        \includegraphics{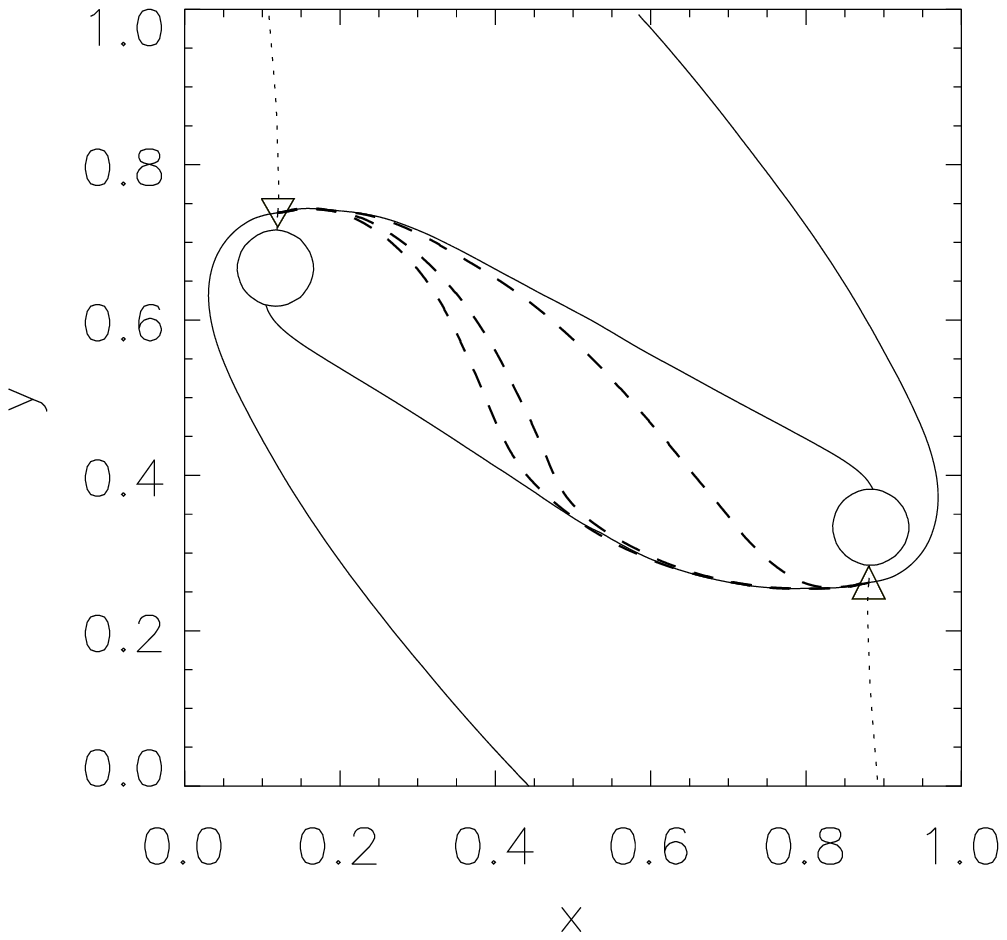}} &
      \resizebox{.30\hsize}{.22\hsize}{%
        \includegraphics{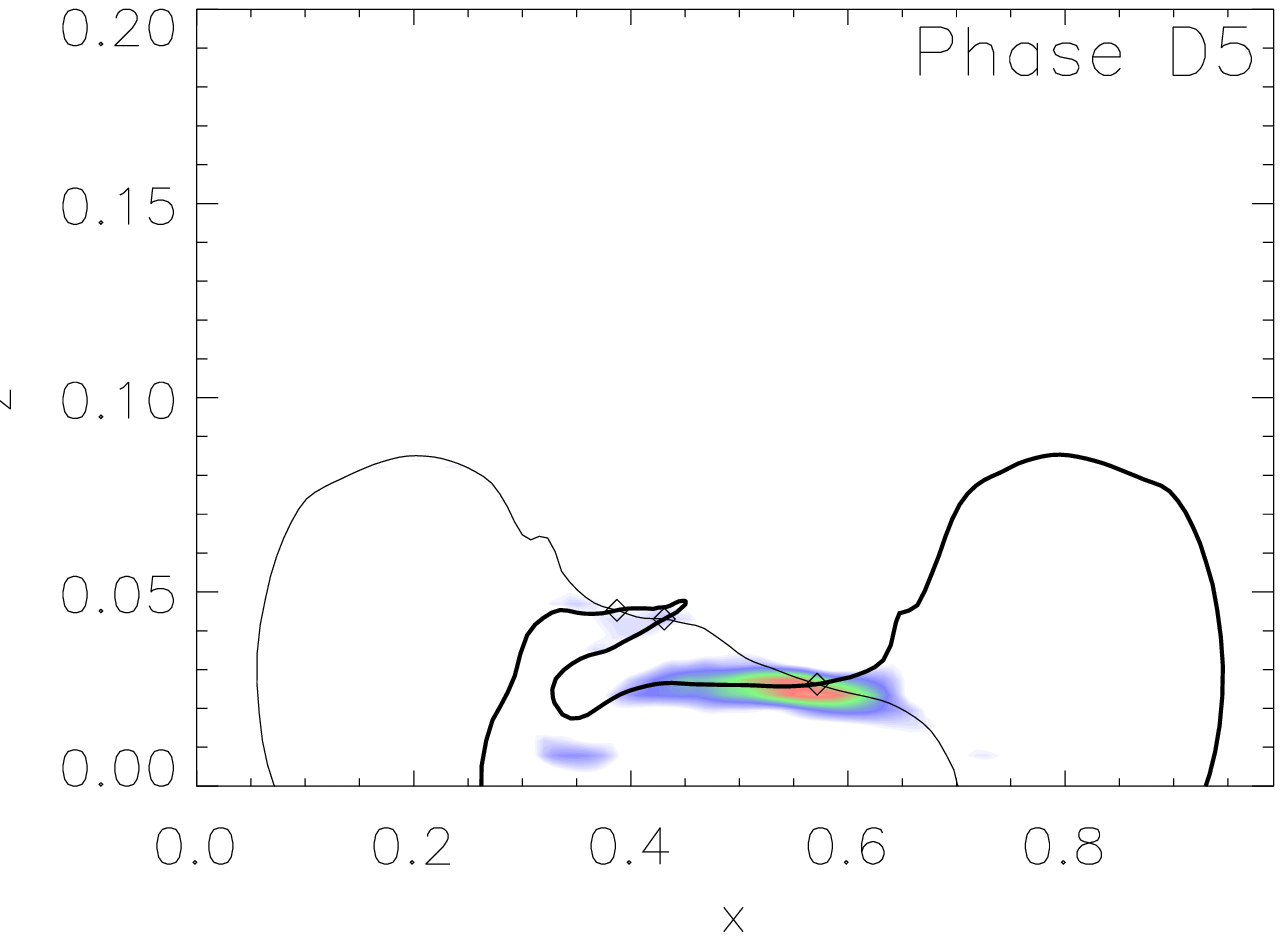}}
  \end{tabular}
\caption{Snapshots of skeletons for the dynamic MHD experiment at $t =
    5.3, 7.6, 14.6, 19.2, 25.8, 40.7$ Alfv\'en times, using the same notation as
    figure~\ref{fig:pot-phases}.  In Column 3, a filled contour plot of current
    intensity is superimposed.
  }
  \label{fig:mhd-phases}
\end{figure}

The two open flux domains break through each other marking the start of this new phase (figure~\ref{fig:mhd-phases}, row 2). Two separators and a closed flux domain are created by way of a global double-separator bifurcation (see \S4 for further 
details).
 The closed flux domain is purely coronal and thus cannot be seen in the 
footprint, though it is 
\clearpage
\clearpage
\begin{table}
  \caption{Characteristics of the dynamical MHD phases}
  \longcaption{The components of each phase for the dynamic MHD experiment based on
    the number of separators ($X$) and the multiplicity of each source pair.  The number in
     brackets denotes the numbers of domains that are coronal for each source pair.  The total number of flux domains (${\cal D}$) is also calculated.}
  \begin{tabular}{*7c} \hline
    Phase & Seps. & \multicolumn{4}{c}{Source Pairs -- Flux Domains} & Total \\
    &  ($X$) & over. & +ve op. & -ve op. & closed & (${\cal D}$) \\ \hline
    D1 & 0 &  1 & 1 & 1 & 0 & 3 \\
    D2 & 2 &  2 & 1 & 1 & 1(1) & 5 \\
    D3 & 1 &  1 & 1 & 1 & 1 & 4 \\
    D4 & 5 &  1 & 3(2) & 3(2) & 1 & 8 \\
    D5 & 3 &  1 & 2(1) & 2(1) & 1 & 6 \\
    D6 & 1 &  1 & 1 & 1 & 1 & 4 \\ \hline
  \end{tabular}
  \label{tab:phases-mhd}
\end{table}
\begin{figure}[ht]  
  \resizebox{7cm}{!}{\includegraphics{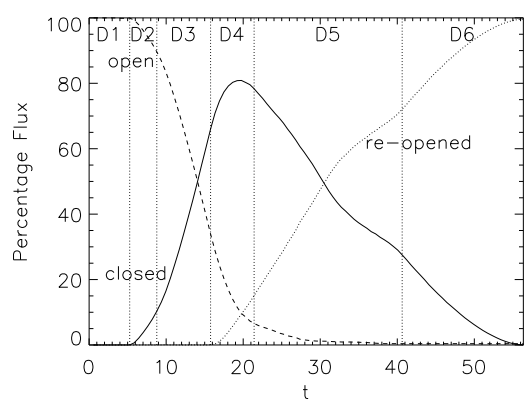}}
  \caption{Percentage of open (dashed), closed (solid) and reopened (dotted)
    flux from the negative source versus the time ($t$) given in Alfv\'en time
    units for the dynamic MHD experiment (from Parnell \& Galsgaard 2004).  Vertical 
    lines denote the transition times between phases.}
  \label{fig:mhd-open}
\end{figure}
\noindent clearly visible in the cross-section. This coronal domain pierces the overlying region creating a flux region of multiplicity two. The field lines in the overlying flux region are no longer simply connected, but form two separate flux domains (table~\ref{tab:phases-mhd}). The spines marked in the footprint denote the lines along which the positive and negative open flux domains fold back and touch the base, thus forming a separatrix surface boundary between the two overlying flux domains (see \S5 for a detailed explanation). 

A current sheet extends along the upper separator.  This is visible
as a tall thin current sheet in the cross-section (figure~\ref{fig:mhd-phases}, row 2, col 3). Here, it is found that separator reconnection is
creating closed magnetic field.  However, around the lower separator reconnection is reopening the field and thus decreasing the flux in the trapped overlying flux domain. The shrinking of the small trapped
overlying domain brings the lower separator to the photosphere signalling a change of topology. 

By the end of this phase 15\% of the flux in the dynamic experiment is closed, whereas in the potential
model all the flux has been closed and some 25\% has reopened again.

\subsubsection*{Phase D3 ($9.68<t<16.96$): Single-Separator Closing}

A global separator bifurcation at the base destroys the lower separator 
and
reduces the overlying field to a simply connected domain once again. 
Thus, phase D3 has just one separator and four flux domains, one of each connectivity (figure~\ref{fig:mhd-phases}, row 3). From figure~\ref{fig:mhd-open} it can be seen that the reduction from two to one 
separator has no effect upon the rate of closing. This is because the 
reconnection at the lower separator was weak.
By the end of this
phase, just 68\% of the flux is closed in this experiment, whereas
all the flux in the potential model has both closed and then reopened, and 
the final phase P4 has begun. 

The separator in this experiment continues to create closed flux, which enlarges the closed flux domain until its encompassing separatrix surfaces expand out to touch the surfaces separating the open and overlying flux domains in two positions above the source plane (figure~\ref{fig:mhd-phases}, row 3, col 3). 
 
\subsubsection*{Phase D4 ($16.96<t<21.14$): Quintuple-Separator Hybrid}

Here, two new separator pairs and four new coronal domains are created (figure~\ref{fig:mhd-phases}, row 4). The separator pairs are created through two global double-separator 
bifurcations. The bifurcations allow the positive 
(negative) open source pair to split into two through the creation of a new 
negative (positive) reopened flux domain generating the four new 
coronal domains. These six open flux domains can be seen in turn in the 
cross-section $y=0.5$ forming an arc around the closed domain (figure~\ref{fig:mhd-phases}, row 4, col 3).
From left to right these domains are: negative open (original), 
positive open (new re-opened), negative open (original), 
positive open (original), negative open (new reopened) and 
positive open (original). Hence, there are two source pairs, each of multiplicity three. Separating these flux 
domains are five separators which all lie on the same pair of separatrix 
surfaces, and hence all connect the same two null points.

The five separators represent five plausible sites for reconnection, but only
three lie in high current regions (figure~\ref{fig:mhd-phases} row 4, col 3); thus significant reconnection is expected only at these sites. We find that reconnection at the central separator is closing the field whilst at the two separators on either side it is reopening the field.  This behaviour is very different from the
potential model where closing and reopening never happen concurrently.  

Reconnection at the lower separators converts the flux in the two small triangular open domains into overlying and closed flux, causing these
separators to fall to the photosphere and the topology to change.

\subsubsection*{Phase D5 ($21.14<t<40.6$): Triple-Separator Hybrid}

A pair of global separator bifurcations destroy the two lower separators, leaving three separators, and reduces the multiplicity of the positive and negative open 
source pairs to two. Thus, here, there are two source pairs of 
multiplicity one, namely: overlying and closed, and two of multiplicity two, namely: positive open and negative open.  
From 
figure~\ref{fig:mhd-open} it is clear that there is still reconnection 
occurring at all the separators; the top separator is still closing the 
field, while the other two separators are reopening the field.  

The driver is switched off at $t=25$. Shortly after this we see that the 
symmetry in the system starts to be lost with the original negative flux being used up much quicker than the original positive flux. Once the original negative domain is emptied of flux this phase ends and a new one beings.

\subsubsection*{Phase D6 ($40.6<t<56.4$): Single-Separator Reopening}
A global double-separator bifurcation again heralds the start of this phase. However, here, we do not see the creation of a separator pair, but rather the loss of the left-hand and middle separators. This then leaves four flux domains, one of each connectivity (figure~\ref{fig:mhd-phases}, row 6). This phase is topologically equivalent to phase P3 of the potential experiment, although here the two separatrix surfaces are rather distorted. Gradually symmetry is restored to the system and what was the right-hand separator now becomes a new central separator. The closed flux continues to be reopened along this separator until almost all the closed flux is used up. 

Just before the positive and negative open flux domains completely separate from one another the experiment ends. However, it is clear that if the experiment had been run for just a short time longer then a final phase D7 would have been entered, in which there would be three simply connected domains and no separators. This stage would have been topologically equivalent to phase D1, but with the positive source and flux domain on the left rather than the right. It would be equivalent to phase P4 from the potential experiment.

\section{Global Bifurcations}

In an individual topological phase the number of reconnection sites remains constant, although their locations may change. A change in topology normally leads to an increase or decrease in the number of reconnection sites and occurs by way of a local or global bifurcation.
During our experiment three global bifurcations are found, two of which are new. The \emph{global separator bifurcation}, in which a single separator 
is formed above the source plane and a mirror separator is formed below, is well known (e.g., Brown \& Priest 1999; Beveridge et al. 2003 and Maclean et al. 2006). During the potential evolution this bifurcation occurs between phases P1 and P2 where it creates a new separator and it also occurs between phases P3 and P4 where is destroys a separator. The bifurcation between phases P2 and P3, however, is new. It is a global bifurcation since the magnetic field in one phase cannot be deformed into another and it does not change the number of nulls or separators. We called it a \emph{global separatrix 
bifurcation}. It changes the handedness or nature of the separatrix surfaces: before this 
bifurcation the positive flux domain is to the right of the negative one, but afterward the order of the flux domains is swapped. Note, that this bifurcation is non-generic since it exists only when the two sources have exactly the same magnitude of flux: for example when the magnitudes are slightly different it will be replaced by two global separator bifurcations, one to an enclosed topology and one away from it.

In the dynamical MHD experiment, we see another new type of global bifurcation in which a 
pair of separators that lie above the source plane is created or destroyed. We call it a
\emph{global double-separator bifurcation}. The separators are created by the
bulging of the two separatrix surfaces through each other. This naturally 
creates two curves of intersection between the two surfaces after they have 
touched and, hence, a pair of separators linking the same pair of nulls. The reverse leads to the destruction of a separator pair.
We suggest that this is one of the most natural ways in which separators can be formed or destroyed in the solar atmosphere. Other possible ways are by a global separator bifurcation, a global spine-fan bifurcation (Maclean et al. 2005) and less commonly by local separator and local double-separator bifurcations.

It is interesting to note that the reconnection that occurs at these pairs of 
separators each creates and destroys different types of flux. For instance, 
in phase D2, the upper separator reconnects positive and negative open 
flux to form closed and overlying flux. However, the lower separator reconnects
closed flux with the overlying flux trapped below this domain to produce more 
positive and negative open flux. And similar opposite reconnection processes 
occur at the two pairs of new separators in phase D4.  This means that for 
the new coronal domain bounded by the pair of separators to grow, the rates 
of reconnection along the separators must be different with the one filling 
the coronal domain dominant. It is natural 
that for a global double-separator bifurcation to destroy two separators the 
dominant reconnection process must be that which destroys the flux in the 
coronal domain bounded by the separator pair.

The reverse reconnection along the new pair of separators is in contrast to 
the behaviour of the separator and mirror separator formed in a global 
separator bifurcation. Here, as one would expect from the name, the 
reconnection along the mirror separator, below the source plane, mirrors the
behaviour along the separator above.  That is their reconnection leads to the
filling the domain they bound which straddles the source plane. Hence, global
separator and global double-separator bifurcations show quite distinct
behaviours.

\section{Multiply-Connected Domains and Magnetic Topology Formulae}
One of the most striking differences between the potential and the dynamical MHD evolution of the magnetic topology is the appearance of multiply-connected source pairs. These are a natural consequence of the global double-separator bifurcation which pierces a hole in a domain, allowing the insertion of a newly created domain, and leading to distinct paths connecting the same pair of sources. Each flux domain of a multiply-connected source pair contains field lines that connect the same pair of sources and forms a simply-connected volume that is bounded by a separatrix surface. 
\begin{center}
\begin{figure}[ht]
    \resizebox{5.25cm}{3.85cm}{\includegraphics{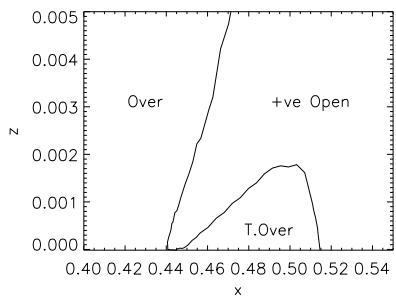}} 
\caption{Cross-section at $y=0.1$ of the magnetic field from the dynamical experiment at $t=7.6$ Alfv\'en times, i.e. during phase D2, showing the positive separatrix surface. Note, $xz$-plane only extends from (0.4,0.0) at the lower left to (0.55,0.005) at the upper right of the region. }
  \label{fig:closeup-D2}
\end{figure}
\end{center}
The global double-separator bifurcation occurs first during the dynamic MHD evolution at the start of phase $D2$ where it traps part of the overlying field whilst creating a new closed domain. Thus, the overlying field region is no longer a simply connected volume, but has two distinct paths, as discussed earlier. From the cross-section in the $y=0.5$ plane (figure~\ref{fig:mhd-phases}, row 2, col 3) it is clear that in this plane the entrapped portion of overlying field is bounded on the left by part of the separatrix surface from the negative null and on the right by part of the surface from the positive null; above, it is edged by a separator. However, what about in the $y=0.1$ plane? Here, clearly, the separatrix surface from the negative null does not cross this plane, but its spine, which cannot divide topologically distinct regions as it is only a line, does. Focusing in on this region (figure~\ref{fig:closeup-D2}) reveals that the positive separatrix surface is deformed and folds over such that it kisses the base at the spine from the negative null. Thus, the region of trapped overlying field is indeed bounded along its entire length by separatrix surfaces: between $y=0$ and the negative null it is enclosed by the positive null's separatrix surface alone; between the two nulls it is bounded by both the positive and negative nulls' separatrix surfaces; and between the positive null and $y=1$ it is bounded by the negative null's separatrix surface alone.

\subsection{Beveridge-Longcope Equation}

Since the global double-separator bifurcation does not occur during the potential evolution of the magnetic field, all the potential source pairs have multiplicity one. Thus, for all the potential phases, the sum over all the source pairs ${\cal D} = \sum_{n=1}^{1} nDn = D_1$. From table~\ref{tab:phases-pot} we see that for every generic phase the difference between ${\cal D}$ and the number of separators ($X$) is 3. This is a consequence of the Beveridge-Longcope equation (equation~\ref{eq:mod-bev-long}), with four sources, including two for the overlying field ($S=4$), and no coronal nulls ($N_c=0$).

In the first phase of the dynamic experiment, D1, has the same topology as phase P1, hence it too must satisfy the Beveridge-Longcope equation. From table~\ref{tab:phases-mhd}, phase D2 has 3 source pairs with multiplicity one and one source pair with multiplicity two, so here ${\cal D}=\sum_{n=1}^{2} nD_n = 1\times3+2\times1=5$. Thus, from equation~\ref{eq:mod-bev-long}, ${\cal D}-X = S-N_c-1=3$, since there are two separators, $X=2$. As in the potential case, of course, $S=4$ and $N_c=0$, since the number of sources has not changed and again no coronal nulls are created.

Phase D3 occurs due to the loss of one separator, which reduces the overlying field to multiplicity one, leading to four simply connected domains and only one separator. Thus, here ${\cal D} = 4$ and ${\cal D}-X=3$, in agreement with equation~\ref{eq:mod-bev-long}.
In phase D4, the global double-separator bifurcation occurs twice. This creates four new separators bringing the total number of separators ($X$) to five. It also creates two source pairs of multiplicity three with two other source pairs of multiplicity one surrounding them. Here, therefore, ${\cal D}=\sum_{n=1}^{3} nD_n = 1\times2+2\times0+3\times2=8$ and once again equation~\ref{eq:mod-bev-long} is satisfied.
To create phase D5 the two source pairs of multiplicity three are reduced to multiplicity two each by the loss of a separator through the source plane. Thus, here ${\cal D}=\sum_{n=1}^{2} nD_n = 1\times2+2\times2=6$, $X=3$ and ${\cal D}-X=3$, so equation~\ref{eq:mod-bev-long} is again true.  

\section{Conclusions}

The magnetic skeleton is found for a magnetic configuration in which two opposite polarity photospheric sources pass by each other in an overlying field. Such an elementary interaction represents a fundamental building block of the coronal heating process. Even though this magnetic field involves just four sources, two of which are at infinity, the topology of the magnetic field is found to evolve through a series of relatively complex states with many multiply-connected source pairs and separators. The evolution found here in the dynamical MHD experiment is not dependent on the form of the resistivity used and has been found in experiments where $\eta$ is uniform or of anomalous/hyper-resistive form and so is likely to be the natural evolution of the field. 

In general, it is expected that in magnetic fields with few sources finding the connectivity of the sources is the same as finding the flux domains, i.e. it is expected that the volumes containing field lines of the same connectivity would be simple and have no holes. However, in our dynamic MHD experiment, we have found three phases (D2,D4 \& D5) where one or more of the source pairs are not simply connected, but have more than one distinct route to link the same two sources. These multiply-connected source pairs arise naturally as a consequence of the global double-separator bifurcation.

This new global double-separator bifurcation, which we have discovered here, is the way in which all the separators found in our dynamic MHD experiment are created. The bifurcation has two interesting consequences: 
\begin{itemize}
\item The new pair of separators engirdle a new coronal domain which is created when two separatrix surfaces touch and reconnection allows them to intersect above the source plane. This is likely to be one of the most natural ways in which separators are formed in the solar atmosphere. The new coronal domain pierces an existing domain making the source pair multiply-connected. Since multiply-connected source pairs are naturally created when a global double-separator bifurcation occurs they are likely to be common.

\item It appears as if the reconnection processes along each separator of a pair  formed by the global double-separator bifurcation do not in general work to fill the same regions. Thus, for the newly formed domain to be created and to grow the reconnection along one separator must be dominant.
\end{itemize}

In the potential evolution of the same magnetic setup, the magnetic field transforms through a non-generic transition state involving an isolated domain surrounded by an infinite number of separators. The dynamic MHD evolution has no such non-generic state, since the global double-separator bifurcation ensures that it is avoided. 

In the potential evolution there is generally only one separator.
In the dynamical MHD experiment there are phases involving 1, 2, 3 or even 5 separators. The reconnection is concentrated around these separators. Thus the separators are the focus for heating and so the detailed spatial distribution of the observed emission from this type of interaction could not be predicted by potential modelling nor, of course, could the temporal distribution of the thermal emission. 

The natural next questions arising from this work are, what is the nature of the reconnection and how rapid is the reconnection. The answer to these questions are not as straightforward as one might initially imagine and will be discussed in a follow-up paper (Parnell et al. 2007). 

\begin{acknowledgements}
ALH would like to thank the University of St Andrews for its financial support during his PhD and CEP acknowledges the support of PPARC through an Advanced Fellowship. KG was supported by the Carlsberg Foundation in the form of a Fellowship. The numerical experiments in this paper were run on the UK MHD Consortium's parallel machine, Copson, funded by SRIF and PPARC. We thank Colin Beveridge and Dana Longcope for useful discussions.
\end{acknowledgements}

\end{document}